\documentclass[aps,onecolumn,preprintnumbers,nofootinbib,superscriptaddress]{revtex4}
\usepackage{eurosym}
\usepackage{amsmath}
\usepackage{graphicx}
\usepackage{amsfonts}
\usepackage{array}
\usepackage{amsthm}
\usepackage{bm}
\usepackage{palatino}
\usepackage{mathpazo}
\usepackage{supertabular}
\usepackage{subfig}
\usepackage[breaklinks]{hyperref}
\usepackage[font=small,labelfont=bf,
justification=justified,
format=plain]{caption}
\usepackage[font=small,labelfont=bf,justification=justified]{caption}
\usepackage{color}
\usepackage{booktabs}
\usepackage{multirow}
\usepackage[labelfont=bf,labelsep=quad,justification=justified]{caption}
\newcommand{\be}{\begin{equation}}
\newcommand{\ee}{\end{equation}}
\newcommand{\ba}{\begin{eqnarray}}
\newcommand{\ea}{\end{eqnarray}}
\newcommand{\bal}{\begin{align}}
\newcommand{\eal}{\end{align}}

\newcommand{\ro}{\rho}

\newcommand{\bw}{\begin{widetext}}
\newcommand{\ew}{\end{widetext}}

\begin{document}

\title{On the possibility of wormhole formation in the galactic halo due to dark matter Bose-Einstein condensates}
\author{Kimet Jusufi}
\email{kimet.jusufi@unite.edu.mk}
\affiliation{Physics Department, State University of Tetovo, Ilinden Street nn, 1200,
Tetovo, North Macedonia}
\affiliation{Institute of Physics, Faculty of Natural Sciences and Mathematics, Ss. Cyril
and Methodius University, Arhimedova 3, 1000 Skopje, North Macedonia}
\author{Mubasher Jamil}
\email{mjamil@zjut.edu.cn}\affiliation{Department of Mathematics, School of Natural Sciences (SNS), National University of Sciences and Technology
(NUST), H-12, Islamabad, Pakistan}
\affiliation{Institute for Astrophysics, Zhejiang University of Technology, Hangzhou, 310032, China}
\author{Muhammad Rizwan}
\email{m.rizwan@sns.nust.edu.pk}
\affiliation{Department of Computer Science,  Faculty of Engineering \& Computer Sciences, National University of Modern Languages, H-9, Islamabad  Pakistan}
\affiliation{Department of Mathematics, School of Natural Sciences, National University of Sciences and Technology, H-12, Islamabad, Pakistan}

\begin{abstract}
It has been recently claimed that dark matter could be in the form of a Bose-Einstein condensate (BEC) in order to explain the dynamics at large distances from the galactic center [Boehmer, Harko, JCAP {\bf 0706}, 025 (2007)]. In this paper we explore the possibility of wormhole formation in galactic halos due to the dark matter BEC. In particular we have found a new wormhole solution supported by BEC dark matter using the expressions for the density profile of the BEC and rotation velocity along with the Einstein field equations to calculate the wormhole red shift function as well as the shape function. To this end, we show that for a specific choose of the central density of the condensates our wormhole solution satisfies the flare our condition.  Furthermore we check the null, weak, and strong condition at the wormhole throat with a radius $r_0$, and shown that in general the energy condition are violated by some arbitrary quantity at the wormhole throat. Using the volume integral quantifier, and choosing reasonable values of parameters we have calculate the amount of BEC exotic matter near the wormhole throat, such that the wormhole extends form $r_0$ to a a cut off radius situated at $`a'$. Moreover we have introduced a Kerr-like metric for a rotating BEC wormhole to study the effect of BEC dark matter on the Lense-Thirring precession frequencies. Namely, we have shown that the obtained precession frequencies lie within a range of typical quasi-periodic oscillations (QPOs).
\end{abstract}
\maketitle



\section{Introduction}

Wormholes are solutions of the Einstein field equations in the context of general theory of relativity. They are discribed as tunnel-like objects which may connect different spacetime regions within the universe, or different universes \cite{Flamm,Einstein}. It is interesting to note that Einstein and Rosen famously proposed a geometric model for elementary particles in terms of the Einstein-Rosen bridge (ERB) which turns out to be unsuccessful \cite{FullerWheeler,whel,Wheeler,Wheeler1,Ellis,Ellis1}.
The study of traversable wormholes (TWs) has attracted a lot of interest in the literature.  Starting from the pioneering work   on traversable wormholes with a phantom scalar field by Ellis \cite{Ellis,Ellis1} and Bronnikov \cite{br1}, followed by other wormhole models studied by Clement \cite{clm}, including the seminal work of Morris and Thorne \cite{Morris}. It has been shown that the geometry of TW requires a spacial kind of exotic matter concentrated at the wormhole throat (to keep the spacetime region open at the throat) implying a special kind of matter which violates the energy conditions, such as the null energy condition (NEC) \cite{Visser1}. It is speculated that such a matter can exists in the context of quantum field theory.  

Dark matter is a hypothetical form of matter that makes up about $27\%$ of the matter-energy composition of the
Universe. Numerous candidates are proposed from particle physics and supersymmetric string theory such as axions and wimps, though, as of today, there is no direct experimental detection
of dark matter. Nevertheless, recent experimental observations strongly
suggest that dark matter reveal its presence in many astrophysical
phenomena, especially in the emergence of
galactic rotation curves \cite{dm1}, the galaxy clusters dynamics \cite{dm2}, to cosmological scales of anisotropies encoded in the cosmic microwave background
measured by PLANCK \cite{dm3}. 
TWs have been studied in different modified theories of gravity, such as $f(R)$ and $f(T)$, where $R$ is the Ricci scalar and $T$ is the torsion of the spacetime, respectively \cite{mj1,mj2,mj3,mj4}, TWs in Born-Infeld gravity \cite{sh1}, TWs with global monopole charge and deflection of light \cite{k1,k2,k3,k4,k5}. In addition to that, wormhole in low dimensional gravity are discussed in \cite{Farooq:2010rh}, wormhole construction in $f(R)$ gravity is studied in \cite{Bahamonde:2016ixz,Rahaman:2013qza}, wormhole construction in teleparallel gravity and $f(T)$ gravity via Noether symmetry is studied in \cite{Bahamonde:2016jqq,Jamil:2012ti}. 

Using the BEC formalism, we adapted the DM density distribution and used to study wormhole construction. In literature, Rahaman et al. \cite{Rahaman1} first proposed the possible existence of wormholes in the outer regions of the galactic halo based on the Navarro-Frenk-White (NFW) density profile, then they used the Universal Rotation Curve (URC) dark matter model to obtain analogous results for the central parts of the halo \cite{Rahaman2}, DM has been adapted as a non-relativistic phenomenon such as NFW profile \& King profile to study WH construction \cite{v2}. In a very interesting idea, Boehmer and Harko, argued that the dark matter required to explain the dynamics of the neutral hydrogen clouds at large distances from the galactic center could be in the form of BEC \cite{harko1}. Cosmological dynamics of BEC dark matter were investigated in \cite{harko2}, gravitational lensing and stability properties of Bose-Einstein condensate dark matter halos have already been studied \cite{harko3}, for the relativistic non-minimally coupled case see \cite{stefano},  dark matter and dark energy from a BEC \cite{Das:2014agf}, dark matter as a mon-relativistic BEC with massive gravitons \cite{Kun:2018ino}, relativistic self-gravitating BEC  with a stiff equation of state \cite{Chavanis:2014hba}, BEC dark matter model at galactic cluster scale \cite{Harko:2015nua,Pires:2012yr,deSouza:2014hwa}. In Ref. \cite{sabin} authors studied the quantum simulation of traversable wormhole spacetimes, while for thin shell BEC see \cite{richarte}. A static and Kerr-like black hole metric due to the BEC dark matter was recently found by \cite{xu}. Motivated by the above works, in the present paper, we study the possibility of TW formation due to non-relativistic dark matter BEC.  

The present paper is organized as follows. In Section II we briefly review the BEC dark matter idea. In Section III,  we construct a new wormhole solution supported by non-relativistic BEC matter. In Section IV, we elaborate the flare-out condition and together with the energy conditions. In Section V, we calculate the exotic matter required to hold these wormholes open. In Section VI, we introduce a Kerr-like metric for a rotating BEC wormhole solution. In Section VII, we study the LT precession of a test gyroscope. In Section VIII, we study the observational aspect by calculating the nodal precession frequencies. Finally in Section V, we comment on our results.

\section{Dark matter as Bose-Einstein condensate}

In this section, we outline the theory of BEC dark matter  following \cite{harko1}. Let us recall that in a quantum system with $N$ interacting condensed bosons, the quantum state of
the bosons can be represented by a single-particle quantum state. The many-body system of interacting bosons in external potential $V_{ext}$ can be described by the Hamiltonian \cite{harko1}
\begin{eqnarray}
\hat{H}&=&\int d\vec{r}\hat{\Psi}^{+}(\vec{r}) \left[ -\frac{%
\hbar ^{2}}{2m}\nabla ^{2}+V_{rot} (\vec{r}) +V_{ext}(\vec{r%
}) \right] \hat{\Psi}(\vec{r}) +
\frac{1}{2}\int d\vec{r}d\vec{r}^{\prime }\hat{\Psi}^{+}(\vec{r}%
) \hat{\Psi}^{+}( \vec{r}^{\prime }) V\left( \vec{r}-\vec{r}%
^{\prime }\right) \hat{\Psi}(\vec{r}) \hat{\Psi}(\vec{r}^{\prime}),  
\end{eqnarray}
in which $\hat{\Psi}(\vec{r}) $ and $\hat{\Psi}^{+}( \vec{r}) $ represent the boson field operators that annihilate and create a particle at the position $\vec{r}$, respectively. Note that $V\left( \vec{r}-\vec{r%
}^{\prime }\right) $ is known as the two-body interatomic potential, and $%
V_{rot}\left( \vec{r}\right) $ stands the potential which encodes the
rotation of the BEC. In the present paper we shall neglect this term, therefore we set $V_{rot}\left( \vec{r}\right)=0$. As for $V_{ext}\left( \vec{r}\right) $, we assume that it is the
gravitational potential $V$, $V_{ext}=V$, and it satisfies the \cite{harko1}
Poisson equation
\begin{align}
      \nabla^{2}V=4\pi G\rho _{m},
\end{align}
where $\rho _{m}=m\rho $ is the mass density inside the
BEC. If we consider only the first
approximation and neglect the rotation of the BEC,
i.e. $V_{rot}=0$, the radius $R$ of the BEC is calculated to be \cite{harko1}
\begin{align}
R=\pi \sqrt{\frac{\hbar ^{2}\alpha}{Gm^{3}}},
\end{align}
where $\alpha$ is called the scattering length, and it is related to the scattering cross section of particles in the condensate. In the literature there are a number of estimations of the mass and scattering length of the condensate dark matter particles. For example, the value of the scattering length $\alpha$, obtained from the astrophysical
observations of the Bullet Cluster was estimated in the range of $10^{-7}$ fm and mass of the dark matter particle 
particle is of the order of $\mu$eV \cite{Harko:2015nua}.  For the density distribution of the dark matter BEC the following result is recovered 
\begin{align}\label{4}
\rho _{DM}\left( r\right) =\rho _{DM}^{(c)}\frac{\sin kr}{kr},
\end{align}
where $k=\sqrt{Gm^{3}/\hbar ^{2}\alpha}$ and $\rho _{DM}^{(c)}$ is the
central density of the condensate, $\rho _{DM}^{(c)}=\rho
_{DM}\left( 0\right) $. The mass profile of the dark condensate
galactic halo $$M_{DM}\left( r\right) =4\pi \int_{0}^{r}\rho
_{DM}\left( r\right) r^{2}dr,$$ yields
\begin{align}
      M_{DM}\left( r\right) =\frac{4\pi \rho _{DM}^{(c)}}{k^{2}}r\left(
      \frac{\sin kr}{kr}-\cos kr\right).
\end{align}

From the last equation one can find the tangential velocity $%
v_{tg}^{2}\left( r\right) =GM_{DM}(r)/r$ of a test particle moving
in the dark halo given by \cite{harko1}
\begin{align}
      v_{tg}^{2}\left( r\right) =\frac{4\pi G\rho _{DM}^{(c)}}{k^{2}}
      \left( \frac{\sin kr}{kr}-\cos kr\right).
      \label{vel}
\end{align}
where $k=\pi/R$.

\section{Traversable Wormholes due to dark matter BEC}

In this section, we shall proceed to find a wormhole solution supported by BEC dark matter Eq. (\ref{4}). To do so, let us consider a static and spherically symmetric spacetime ansatz, commonly termed as Morris-Thorne traversable wormhole, which in Schwarzschild coordinates can be written as follows (from now on we shall use the natural units $G=c=\hbar=1$)
\begin{equation}
\mathrm{d}s^{2}=-e^{2\Phi (r)}\mathrm{d}t^{2}+\frac{\mathrm{d}r^{2}}{1-\frac{b(r)}{r}}+r^{2}\left(
\mathrm{d}\theta ^{2}+\sin ^{2}\theta \mathrm{d}\varphi ^{2}\right),  \label{5}
\end{equation}
in which $\Phi (r)$ and $b(r)$ are known as the redshift and shape
functions, respectively. Of course, in the wormhole geometry the redshift
function $\Phi (r)$ should be finite in order to avoid the
formation of an event horizon and should tend to zero for large $r$ to ensure asymptotic flatness. On the other hand the shape function $b(r)$ determines the wormhole geometry, with the following condition $b(r_{0})=r_{0}$, where $r_{0}$ is the radius of the wormhole throat. Consequently, it follows that the shape function must satisfy the flaring-out condition: 
\begin{equation}
\frac{b(r)-rb^{\prime }(r)}{b^{2}(r)}>0, 
\end{equation}%
in which $b^{\prime }(r)=\frac{db}{dr}<1,$ must hold at the throat of the
wormhole. We recall that the other hand Einstein's field equations  reads
\begin{equation}
G_{\mu \nu }\equiv R_{\mu \nu }-\frac{1}{2}g_{\mu \nu }R=8\pi  \mathcal{T}_{\mu \nu },
\label{68}
\end{equation}%
where $\mathcal{T}_{\mu \nu }$ is the stress-energy-momentum tensor describing the distribution and charactaristics of matter via energy density and pressure. For the BEC dark matter, we shall employ an anisotropic fluid with the following energy-momentum tensor components 
\begin{equation}
 {\mathcal{T}^{\mu}}_{ \nu }=\text{diag}\left( -\rho ,\mathcal{P}_{r},\mathcal{P}_{\theta },%
\mathcal{P}_{\varphi }\right).  \label{16}
\end{equation}%

The components of Einstein tensor for a generic wormhole metric \eqref{5} are computed as follows
\begin{eqnarray}
G_{t}^{t} &=&-\frac{b^{\prime }(r)}{r^{2}},  \\
G_{r}^{r} &=&-\frac{b(r)}{r^{3}}+2\left( 1-\frac{b(r)}{r}\right) \frac{\Phi
^{\prime }}{r},   \\
G_{\theta }^{{\theta }} &=&\left( 1-\frac{b(r)}{r}\right) \Big[\Phi ^{\prime
\prime }+(\Phi ^{\prime })^{2}-\frac{b^{\prime }r-b}{2r(r-b)}\Phi ^{\prime }-\frac{b^{\prime }r-b}{2r^{2}(r-b)}+\frac{\Phi ^{\prime }}{r%
}\Big],   \\
G_{\varphi }^{{\varphi }} &=&G_{\theta }^{{\theta }}.  \label{73n}
\end{eqnarray}

Using these equations the energy-momentum components yields
\begin{eqnarray}
\rho (r) &=&\frac{b^{\prime }(r) }{8\pi  r^{2}} ,   \\
\mathcal{P}_{r}(r) &=&\frac{1}{8\pi  }\left[ 2\left( 1-\frac{b(r)}{r}\right) 
\frac{\Phi ^{\prime }}{r}-\frac{b(r)}{r^{3}}\right]
, \\
\mathcal{P}(r) &=&\frac{1}{8\pi  }\left( 1-\frac{b(r)}{r}\right) \Big[\Phi
^{\prime \prime }+(\Phi ^{\prime })^{2}-\frac{b^{\prime }r-b}{2r(r-b)}\Phi
^{\prime }- \frac{b^{\prime }r-b}{2r^{2}(r-b)}+\frac{\Phi ^{\prime }}{r%
}\Big].  \label{18}
\end{eqnarray}%
with $\mathcal{P}=\mathcal{P}_{\theta }=\mathcal{P}_{\varphi }$. 

Let us recall that the rotational velocity of a test particle in spherically symmetric space-time, within the equatorial
plane is determined by \cite{harko1}
\begin{equation}\label{vtg}
v_{tg}^2(r)=r \,\Phi'(r).
\end{equation}

Combining Eq. (\ref{vtg}) with the expression for the test particle moving in the dark halo given by Eq. (6), we find
\begin{align}
   r \,\Phi'(r) =\frac{4\,\rho _{DM}^{(c)}R^2}{\pi}
      \left[ \frac{\sin  \left(\frac{\pi r}{R}\right)}{ \frac{ \pi r}{R}}-\cos\left(\frac{\pi r}{R}\right)\right].
\end{align}

\begin{figure}[h!]
\includegraphics[width=0.50\textwidth]{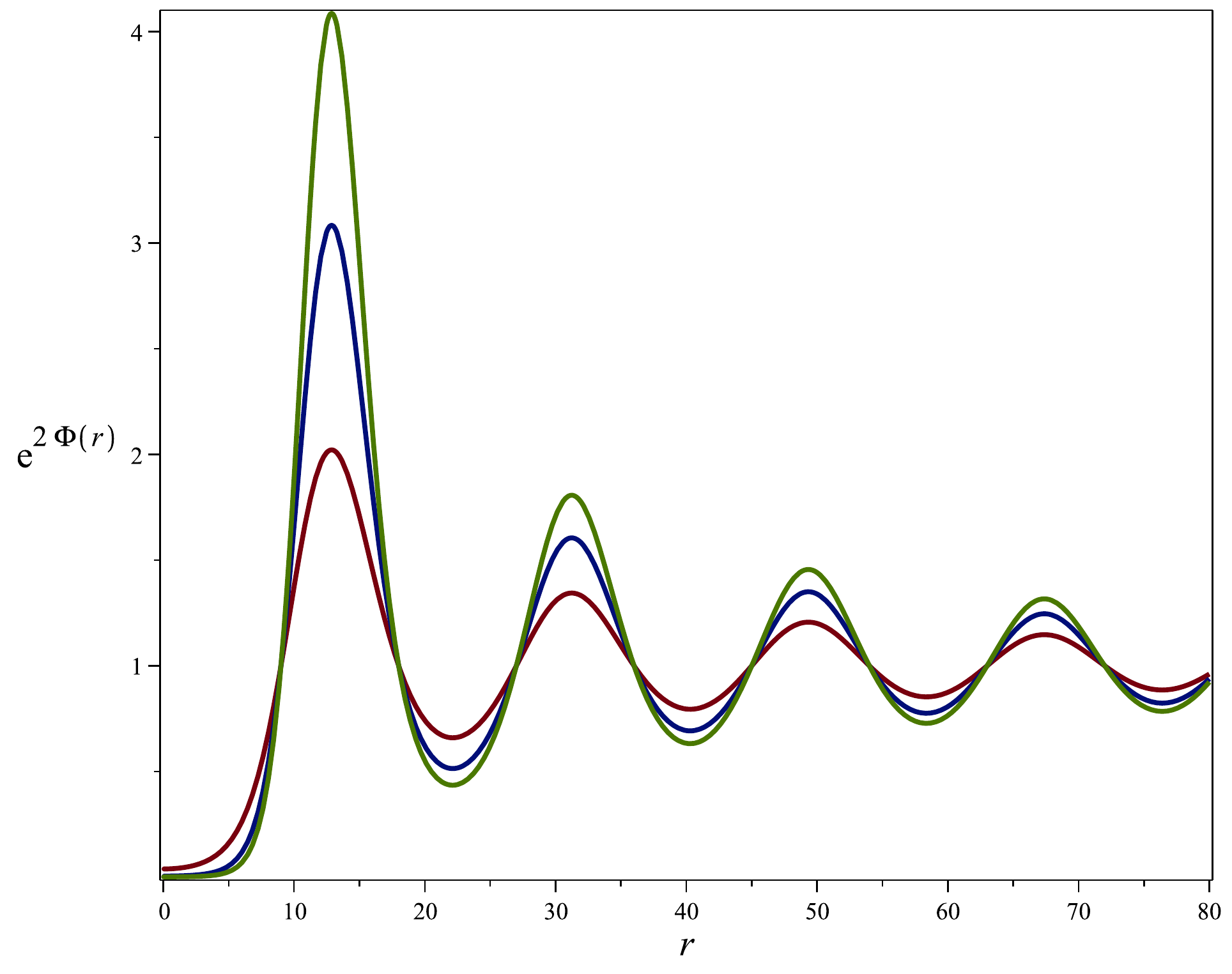}
\includegraphics[width=0.45\textwidth]{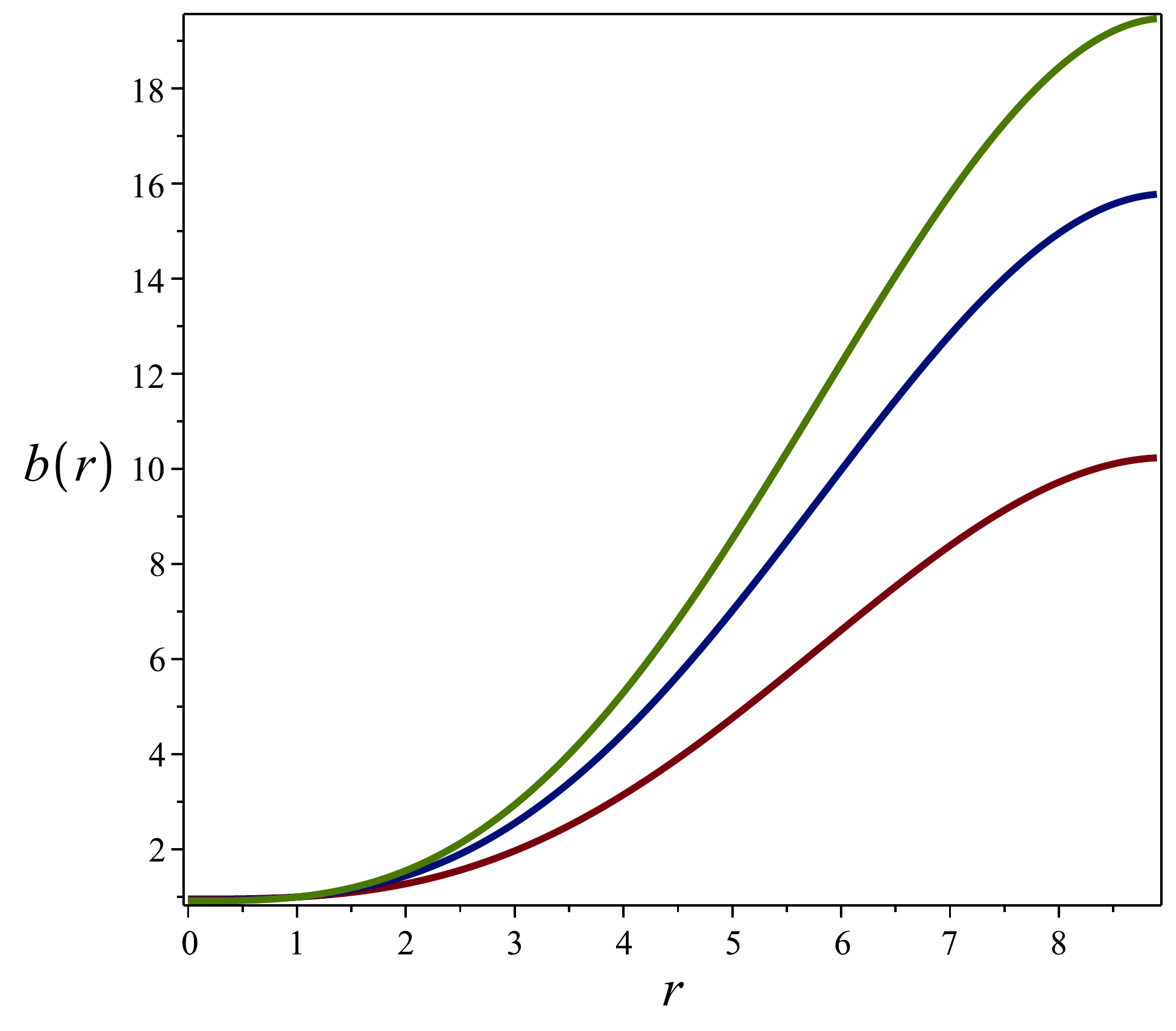}
\caption{{\protect\small \textit{Left panel: It is shown the behavior of 
$\exp(2\Phi(r))$ as a function of $r$. Right panel: We plot the behavior of shape function
$b(r)$ as a function of $r$. In both plots we have chosen $R=9$ in units of kpc, and $\rho_{DM}^{c}=0.005$ (red), $\rho_{DM}^{c}=0.008$ (blue), $\rho_{DM}^{c}=0.01$ (green), in the units of $M_{DM}$ halo/kpc$^3$.}}}
\label{f1}
\end{figure}

Solving this differential equation we find
\begin{align}
   \Phi(r)=C_1-\frac{4\,\rho _{DM}^{(c)} R^3}{\pi^2 r}\sin\left( \frac{ \pi r}{R}\right),
\end{align}
where $C_1$ is a constant of integration. Thus, the $g_{tt}$ component of the metric tensor yields 
\begin{align}
  \exp\left(2 \Phi(r)\right)=\mathcal{D}\exp\left[-\frac{8\,\rho _{DM}^{(c)} R^3}{\pi^2 r}\sin\left( \frac{ \pi r}{R}\right)\right].
\end{align}

However the new constant $\mathcal{D}$ is absorbed by rescaling the time coordinate $t \to \mathcal{D}t$. Note that the function $\exp\left(2 \Phi(r)\right)$ is constrained by the relation
\begin{equation}
  \lim_{r \to R}\exp\left(2 \Phi(r)\right)= \lim_{r \to R}\left\lbrace \exp\left[-\frac{8\,\rho _{DM}^{(c)} R^3}{\pi^2 r}\sin\left( \frac{ \pi r}{R}\right)\right]\right\rbrace=1.
\end{equation}

In fact we can see this graphically from the Fig. 1. Next, we assume that the density of the wormhole matter to be the profile density of the BEC, i.e.  $\rho(r)=
\rho _{DM}\left( r\right) $.  This can be justified from the fact that the velocity of test particle considered in this model is much smaller than the speed of light, hence one can approximate the density profile with the energy density in the Einstein field equations. From Eq. (15) it follows that
\begin{align}
   b'(r)=8  r  \rho _{DM}^{(c)}R\sin \left( \frac{ \pi r}{R}\right).
\end{align}
Solving this differential equation we find the solution
\begin{align}
   b(r)=C_2+\frac{8\,G \rho _{DM}^{(c)} R^3}{\pi^2}\left[ \sin \left( \frac{ \pi r}{R}\right)-\cos \left( \frac{ \pi r}{R} \right)\frac{\pi r}{R}  \right].
\end{align}

Using the condition $b(r_0)=r_0$, we can fix the constant $C_2$ for the wormhole shape function. The final result for the shape function is obtained as follows
\begin{eqnarray}
     b(r)&=&r_0+\frac{8\, \rho _{DM}^{(c)} R^3}{\pi^2}\left[\sin \left( \frac{ \pi r}{R}\right)-\sin \left( \frac{ \pi r_0}{R}\right)\right]- \frac{8\, \rho _{DM}^{(c)} R^2}{\pi}\left[\cos \left( \frac{ \pi r}{R}\right)r-\cos \left( \frac{ \pi r_0}{R}\right)r_0\right].
\end{eqnarray}

In a similar way, the functions $b(r)/r$ is constrained by the following relation
\begin{eqnarray}
  \lim_{r \to R} \frac{b(r)}{r}&=&\frac{r_0}{R}+\frac{8\rho _{DM}^{(c)}R}{\pi^2}\left[\cos \left( \frac{ \pi r_0}{R}\right) \pi r_0 - \sin \left( \frac{ \pi r_0}{R}\right)   \right]+\frac{8\rho _{DM}^{(c)}R^2}{\pi}.
\end{eqnarray}

In this sense, our solution is not asymptotically flat. Finally the wormhole line element supported by BEC matter can be recast in the following form
\begin{align}
ds^2= -\exp\left[-\frac{8\,\rho _{DM}^{(c)} R^3}{\pi^2 r}\sin\left( \frac{ \pi r}{R}\right)\right]dt^2+\frac{dr^2}{1-\frac{b(r)}{r}}++r^{2}\left(
\mathrm{d}\theta ^{2}+\sin ^{2}\theta \mathrm{d}\varphi ^{2}\right).
\end{align}

In order to maintain the wormhole structure, the flaring out condition needs to be satisfied in order to keep the wormhole mouth open. This condition at the wormhole throat region is given by the following relation (by restoring temporary the constants $G$ and $c$): 
\begin{align}
   b'(r_0)=\frac{8 G r_0  \rho _{DM}^{(c)}R}{c^2}\sin \left( \frac{ \pi r_0}{R}\right)<1.
\end{align}

Taking $r_0=9.11$ kpc for the wormhole throat radius,  central density for BEC $\rho_{DM}^{c}=2\times 10^{-25}$ g/cm$^3$ \cite{harko1}, BEC radius $R=16$ kpc, we find that indeed this condition is indeed satisfied
\begin{align}
   b'(r_0)=\frac{8 G r_0  \rho _{DM}^{(c)}R}{c^2}\sin \left( \frac{ \pi r_0}{R}\right)=1.60 \times \,10^{-7}<1.
\end{align}

\begin{figure}[h!]
\center
\includegraphics[width=0.54\textwidth]{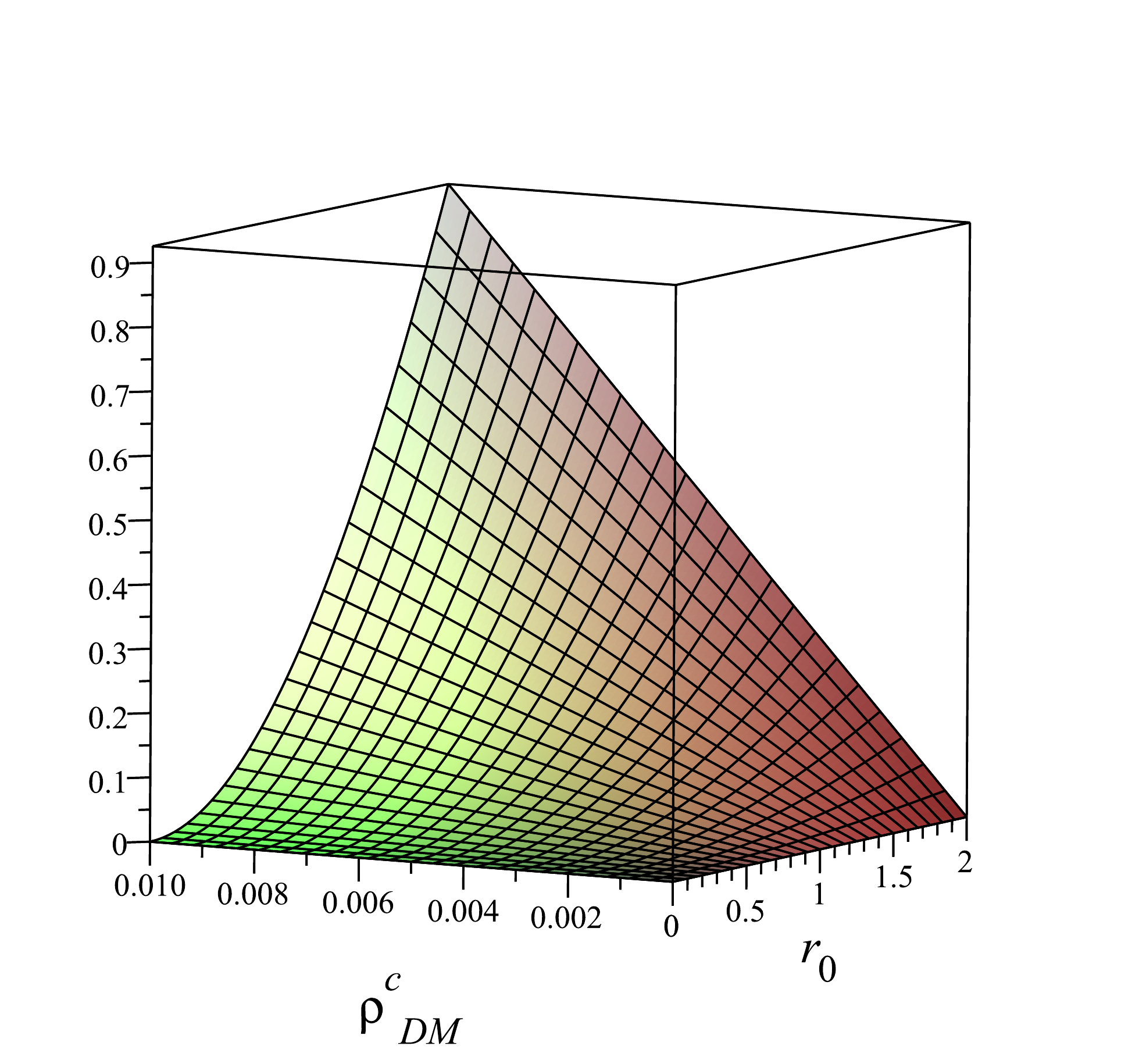}
\caption{{\protect\small \textit{The figure shows the behavior of 
$b'(r_0)$ as a function of $r_0$ and $\rho_{DM}^{c}$, for chosen $R=9$ and $\rho \in [0, 0.01]$ and $r_0 \in (0, 2]$ in units $G=c=\hbar=1$. We note that only $r_0>0$ is physically acceptable.}}}
\end{figure}

For more information, in Fig. 2 we plot a domain of parameters which satisfies the flare-out condition. Since the wormhole metric is not asymptotically flat, we can employ matching conditions by truncating the wormhole metric at radius $a$ and connecting with the exterior Schwarzschild black hole metric, as the later corresponds to vacuum solution and asymptotically flat. Now imposing the continuity we find \cite{raha}
\begin{eqnarray}
\exp\left[-\frac{8\,\rho _{DM}^{(c)} R^3}{\pi^2 a}\sin\left( \frac{ \pi a}{R}\right)\right]=1-\frac{2M}{a},\,\,\,\,  1-\frac{b(a)}{a}=1-\frac{2M}{a}.
\end{eqnarray}

From the second equation we find $b(a)=2M$.  Solving these equations implicitly provide the value of truncated radius $a$, where  the  matching occurs.  Note that in the present paper $\rho_{DM}^{c}$ has units of $M_{DM}$ halo/kpc$^3$, $r$ has units of $M_{DM}$ halo, while $R$ and $r_0$ have units of kpc.

\section{Energy conditions}

 Given the redshift function and the shape function, we can compute the energy-momentum components. In particular for the radial component we find
\begin{eqnarray}\notag
 \mathcal{P}_r(r)&=&\frac{1}{8 \pi^5 r^4}\Big[  (\rho _{DM}^{(c)})^2\cos^2 \left( \frac{ \pi r}{R}\right)(64 R^6-64 R^4 \pi^2 r^2)+ 64 r R^2 \pi \rho _{DM}^{(c)}\cos \left( \frac{ \pi r}{R}\right)\left\lbrace \zeta_1+2\sin \left( \frac{ \pi r}{R}\right)\rho _{DM}^{(c)}  \right\rbrace \\
 &-& 64 R^3 \sin \left( \frac{ \pi r}{R}\right)\rho _{DM}^{(c)}\zeta_1 -8 \pi^2 \rho _{DM}^{(c)}r R^2   \left\lbrace \cos \left( \frac{ \pi r_0}{R}\right)\pi r_0 -\sin \left( \frac{ \pi r_0}{R}\right) R \right\rbrace- 64 R^6 (\rho _{DM}^{(c)})^2-\pi^4 r r_0 \Big],
\end{eqnarray}
where 
\begin{align}
\zeta_1=\rho _{DM}^{(c)}\cos \left( \frac{ \pi r_0}{R}\right) \pi R^2 r_0 - R^3 \sin \left( \frac{ \pi r_0}{R}\right)\rho _{DM}^{(c)}+\frac{r_0 \pi^2}{8}.
\end{align}

On the other hand using Eq. (18) for the tangential component of the pressure we find the following result
\begin{widetext}
\begin{eqnarray}\notag
 \mathcal{P}(r)&=&\frac{1}{16 \pi^7 r^5}\Big[  -256(\rho _{DM}^{(c)})^3 R^9 \sin^3 \left( \frac{ \pi r}{R}\right) +256 R^4 (\rho _{DM}^{(c)})^2 R^9 \sin^2 \left( \frac{ \pi r}{R}\right)\\
 &\times & \left\lbrace 3R^4 r \pi \rho _{DM}^{(c)} \cos \left( \frac{ \pi r}{R}\right) + \frac{\pi^2}{2}\left( R^2(r-\frac{r_0}{4})-\frac{3 r^3 \pi^2}{4} \right)+  \mathcal{A} \right\rbrace - 512 \pi R r \rho _{DM}^{(c)}  \sin \left( \frac{ \pi r}{R}\right)\\\notag
 &\times & \left\lbrace  \frac{3 R^6 r \pi (\rho _{DM}^{(c)})^2  \cos^2 \left( \frac{ \pi r}{R}\right)}{2}+R^2 \rho _{DM}^{(c)}\cos \left( \frac{ \pi r}{R}\right)  \left[\mathcal{A}+\frac{\pi^2}{2}\left( R^2(r-\frac{r_0}{4})-\frac{3 r^3 \pi^2}{8} \right)\right]+ \frac{3 \mathcal{B} \pi}{16} (R^2-\frac{2 r^2 \pi^2}{3}) \right\rbrace \\\notag
 &+& 256 r^2 \pi^2 \left\lbrace R^6 r \pi (\rho _{DM}^{(c)})^3 \cos^3 \left( \frac{ \pi r}{R}\right)+R^4 (\mathcal{B}    +\frac{r \pi^2 }{2}) (\rho _{DM}^{(c)})^2 \cos^2 \left( \frac{ \pi r}{R}\right) + \frac{3 R^2 \mathcal{B}}{8}\rho _{DM}^{(c)}\cos \left( \frac{ \pi r}{R}\right)\right\rbrace - \frac{\mathcal{B}\pi^2}{32} \Big],
\end{eqnarray}
\end{widetext}
where 
\begin{equation}
\mathcal{A}= R^5 \rho _{DM}^{(c)}\sin \left( \frac{ \pi r_0}{R}\right)-R^4 \pi r_0 \rho _{DM}^{(c)}\cos\left( \frac{ \pi r_0}{R}\right),
\end{equation}
\begin{equation}
\mathcal{B} = - \pi R^2 r_0 \rho _{DM}^{(c)}\cos \left( \frac{ \pi r_0}{R}\right)+\rho _{DM}^{(c)}\sin \left( \frac{ \pi r_0}{R}\right)R^3-\frac{r_0 \pi^2}{8}.
\end{equation}

\begin{figure}[h!]
\includegraphics[width=0.44\textwidth]{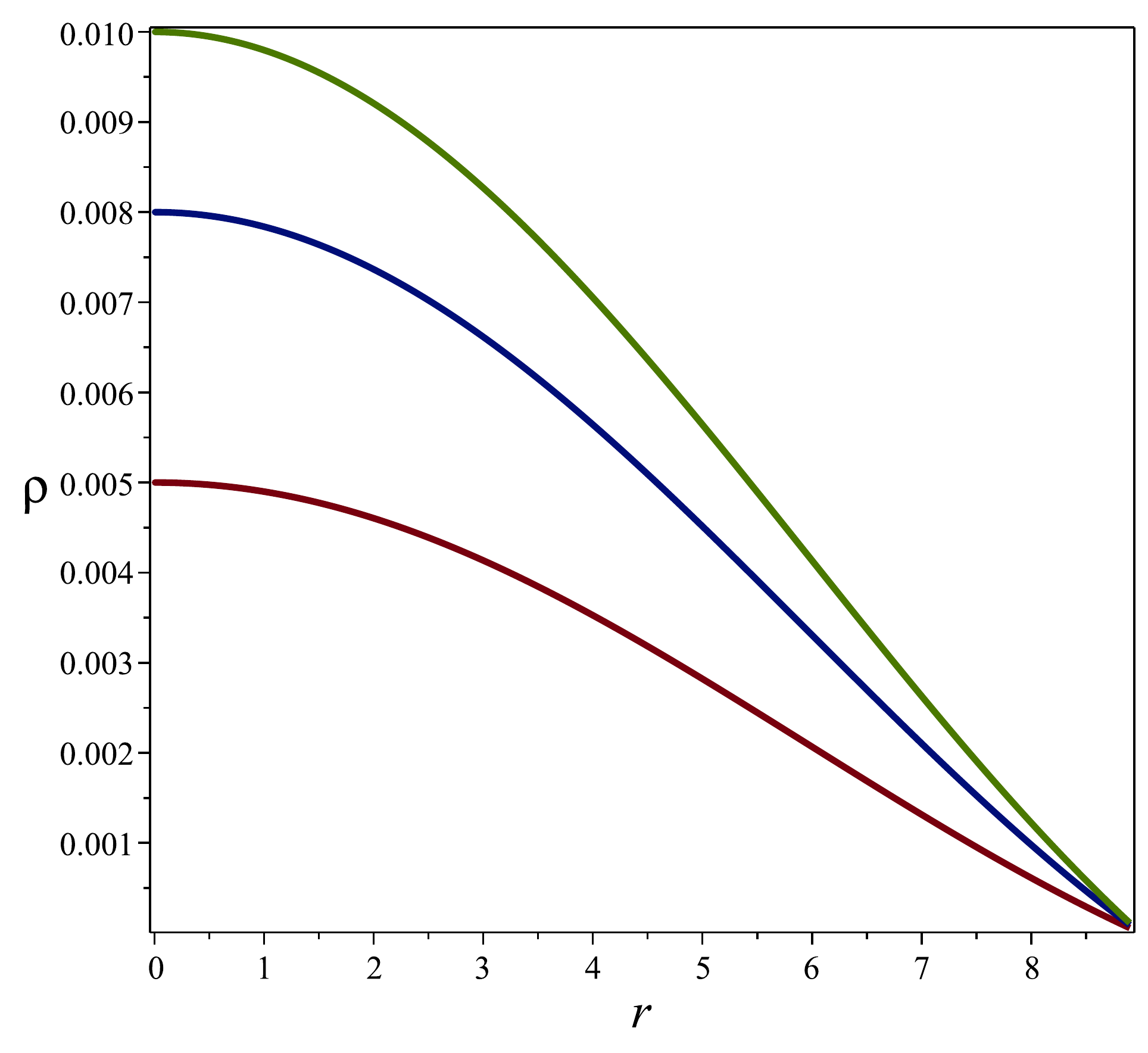}
\includegraphics[width=0.51\textwidth]{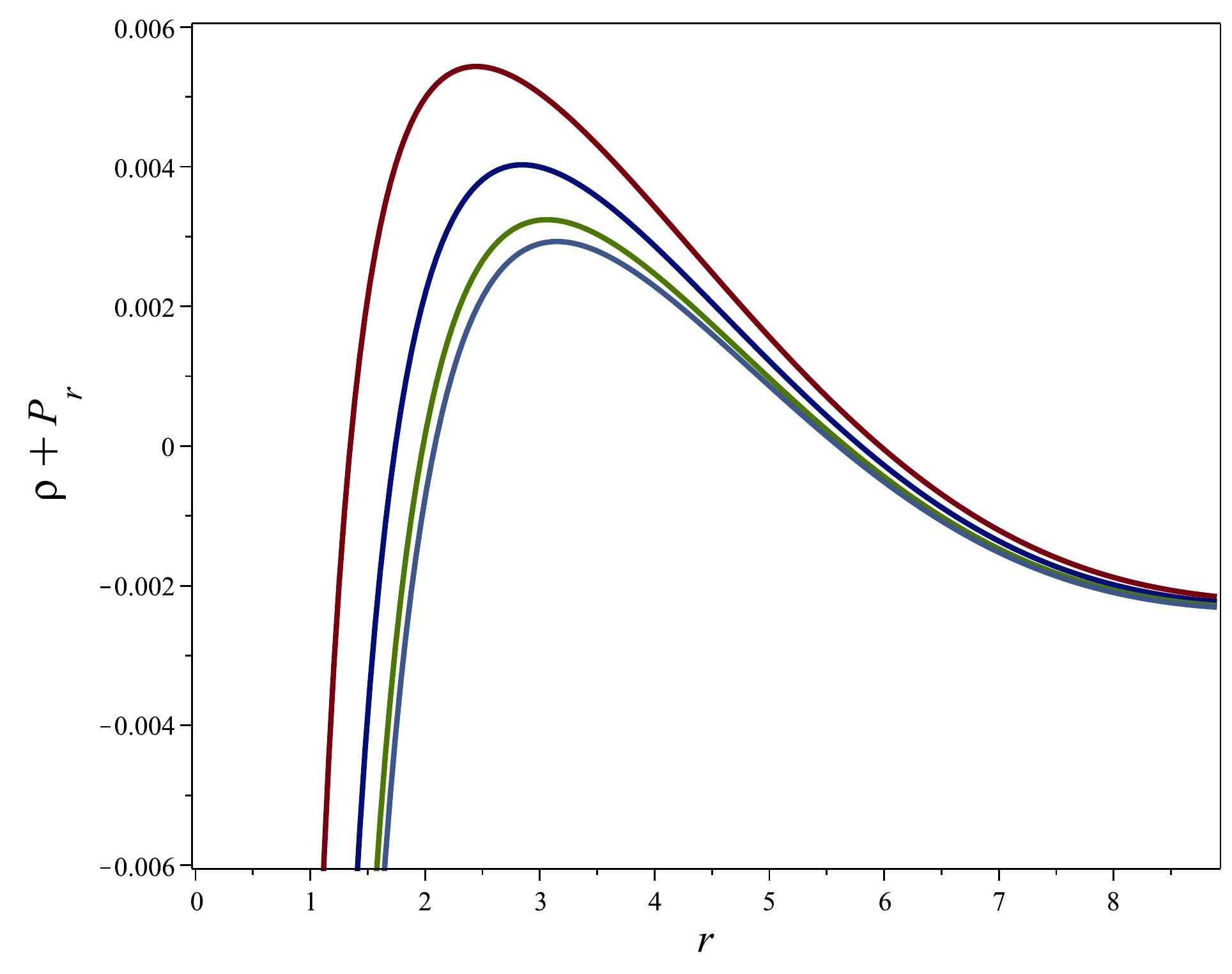}
\caption{{\protect\small \textit{Left panel: The figure shows the variation of the
$\rho_{DM}$ as a function of $r$ with $R=9$, for chosen $\rho_{DM}^{c}=0.005$ (red), $\rho_{DM}^{c}=0.008$ (blue), $\rho_{DM}^{c}=0.01$ (green). Right panel: It is shown the behavior of $\rho_{DM}+P_r$ as a function of $r$ for chosen $R=9$, and $\rho_{DM}^{c}=0.01$. Note that, $r_0=0.5$, $r_0=1$, $r_0=1.5$ and $r_0=2$, from left to right, respectively.}}}
\end{figure}

Next let use discuss the issue of energy conditions and make some
regional plots to check the validity of all energy conditions. Recall that the WEC is
defined by $T_{\mu \nu }U^{\mu }U^{\nu }\geq 0$ i.e.,
\begin{equation}
\rho \geq 0, \,\,\,\,\,\,\,\,\,\,\, \rho (r)+\mathcal{P}_{r}(r)\geq 0,
\end{equation}
where $T_{\mu \nu }$ is the energy momentum tensor and $U^{\mu }$ denotes the timelike vector. This means that
local energy density is positive and it gives rise to the continuity of NEC,
which is defined by $T_{\mu \nu }k^{\mu }k^{\nu }\geq 0$ i.e., 
\begin{equation}
\rho (r)+\mathcal{P}_{r}(r)\geq 0, 
\end{equation}
where $k^{\mu }$ is a null vector. The strong energy condition (SEC)  stipulates that
\begin{equation}
\rho (r)+2\mathcal{P}(r)\geq 0,
\end{equation}
and
\begin{equation}
\rho (r)+\mathcal{P}_{r}(r)+2\mathcal{P}(r)\geq 0.
\end{equation}

\begin{figure}[h!]
\includegraphics[width=0.46\textwidth]{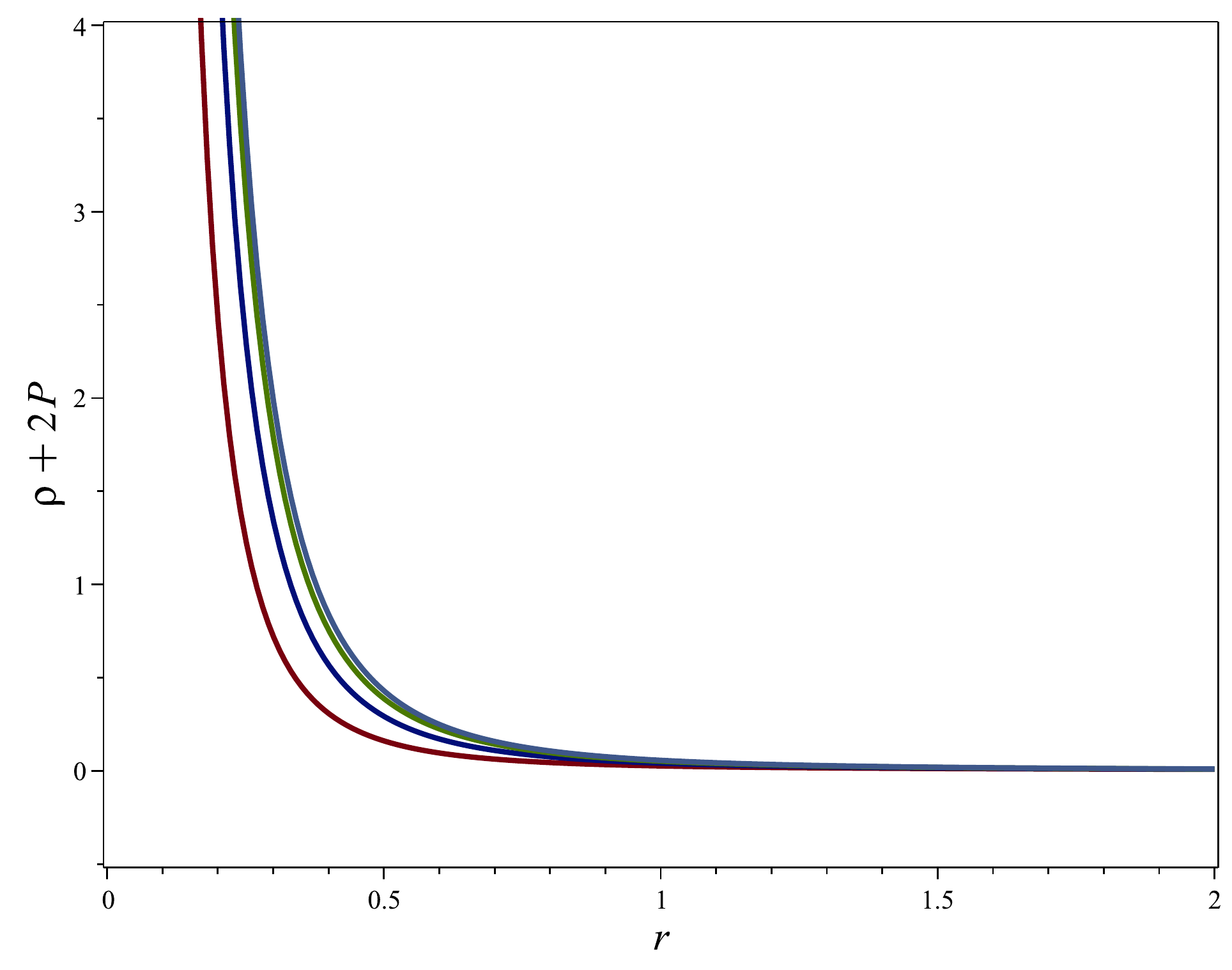}
\includegraphics[width=0.52\textwidth]{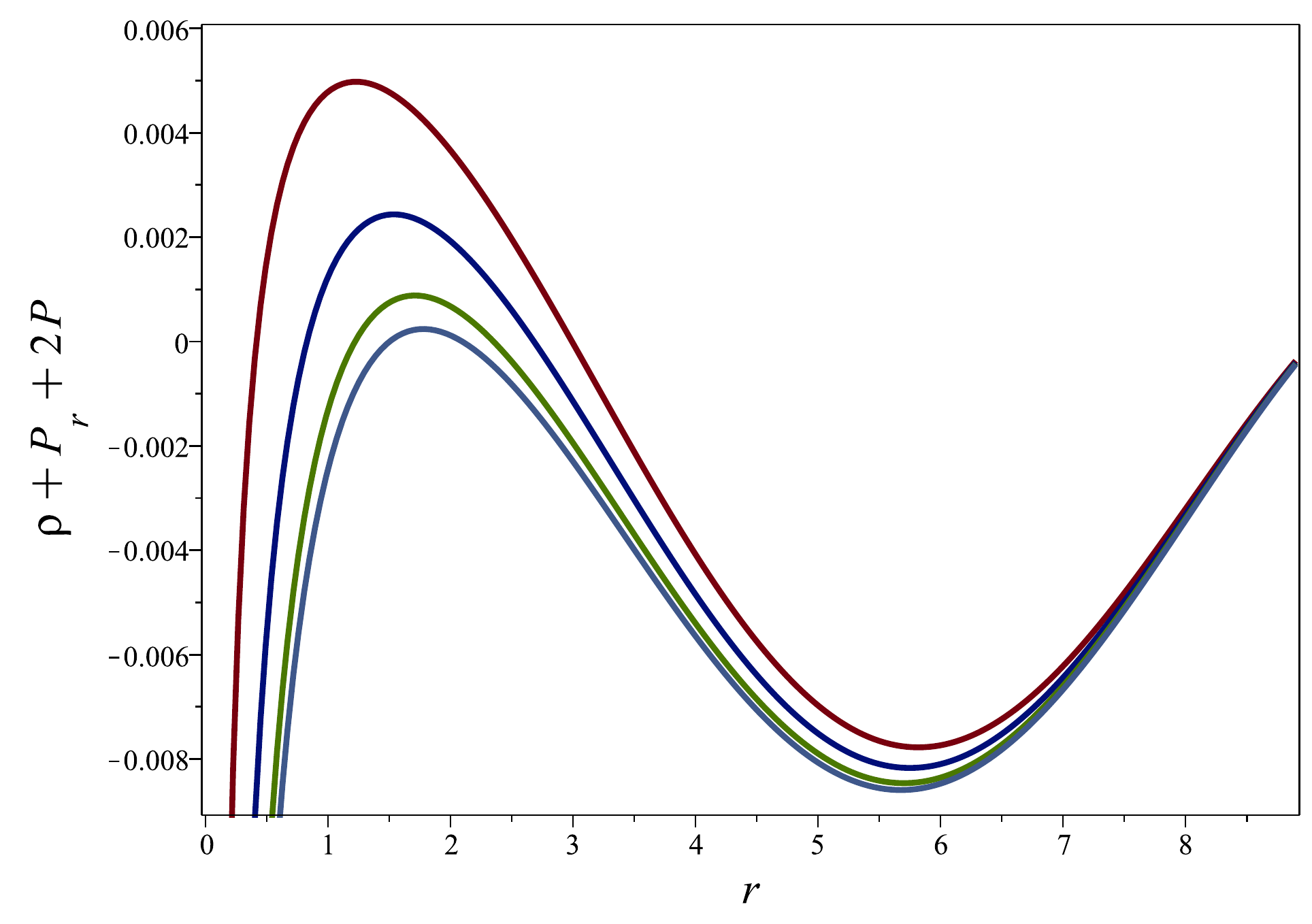}
\caption{{\protect\small \textit{ Left panel: It is shown the variation of 
$\rho_{DM}+2P$ as a function of $r$, with $R=9$ and $\rho_{DM}^{c}=0.01$. Right panel: It is shown the the variation of $\rho_{DM}+P_r+2P$ as a function of $r$, with $R=9$ and $\rho_{DM}^{c}=0.01$. Note that in both plots we have chosen, $r_0=0.5$, $r_0=1$, $r_0=1.5$ and $r_0=2$, from left to right, respectively.}}}
\end{figure}

At first sight from Figs. 3-4 looks like there exists a domain of parameters where the NEC, WEC, and SEC, are satisfied for reasonable values of $r$. In all plots we have chosen values for the wormhole throat; $r_0=\{0.5,1,1.5,2\}$ with $\rho_{DM}^{c}=0.01$, such that the flare out condition is satisfied, namely $b'(r_0)<1$ in all cases. A careful numerical analyses shows that $\left(\rho+\mathcal{P}_r+2P\right)\vert_{r_0=\{0.5;1;1.5;2\}}=\{0.00155,0.00124,0.00074, 0.00011\}>0$ at the wormhole throat, $r=r_0$.  Unfortunately, we find that $\left(\rho+{\mathcal{P}_r}\right)\vert_{r_0=\{0.5;1;1.5;2\}}=\{-1.14920,-0.02999,-0.00813,-0.00073 \}<0$ by a very small quantity. This shows that the energy condition in general is violated by a very small and arbitrary quantity at the wormhole throat. From a quantum mechanical point of view, it is known that quantum fluctuations violate most energy conditions without any restrictions and this opens the possibility that quantum fluctuations may play an important role in the wormhole stability.  On the other hand, by choosing $\rho_{DM}^c>0.01$, say in the interval $\rho_{DM}^c=[0.01,0.05]$, all energy conditions holds, unfortunately, in that case, the flare out condition can not be satisfied. Furthermore we note that only $r_0>0$ is physically acceptable in all plots.

\section{Amount of exotic matter}

An interesting quantity to consider is the ``volume integral quantifier,'' which basically quantifies the amount of exotic matter required for wormhole maintenance. This quantity is related only to $\rho$  and $\mathcal{P}_r$, not to the transverse components, and is defined in terms of the following definite integral \cite{viser1}
\begin{eqnarray}
I_V=\oint [\rho+\mathcal{P}_r]~\mathrm{d}V=2 \int_{r_0}^{\infty} \left(  \rho_{DM}+\mathcal{P}_r\right)~\mathrm{d}V,  
\end{eqnarray}
which can be written also as
\begin{eqnarray}
I_V =8 \pi \int_{r_0}^{\infty} \left(  \rho_{DM}+\mathcal{P}_r  \right)r^2  dr.
\end{eqnarray}

As we already pointed out, the value of this volume-integral encodes information about the ``total amount"
of exotic matter in the spacetime, and we are going to evaluate this integral for 
our shape function $b(r)$. It is convenient to introduce a cut off such that the wormhole extends form $r_0$ to a radius situated at $`a'$ and then  we get the very simple result
\begin{equation}
\mathcal{I}_V=    8 \pi \int_{r_0}^{a} \left(  \rho_{DM}+\mathcal{P}_r  \right)r^2  dr.
\end{equation}

In the special case $a \rightarrow r_0$, we should find $\int{(\rho+\mathcal{P}_r)} \rightarrow 0$. Such a wormhole is supported by arbitrarily small quantities of violating matter. Evaluating the above integral we find that
\begin{eqnarray}\notag
\mathcal{I}_V&=& \frac{1}{\pi^4 a}\Big[-32 R^6  (\rho _{DM}^{(c)})^2\cos(\frac{2 \pi a}{R})- 32 \rho _{DM}^{(c)} \pi R^2 \cos(\frac{\pi r_0}{R}) \left\lbrace R^3 a \rho _{DM}^{(c)} \sin(\frac{\pi r_0}{R}) -2r_0 \Xi_1 \right\rbrace \\\notag
&-& 64 \rho _{DM}^{(c)} R^3 \sin(\frac{\pi r_0}{R})\Xi_2 - 32 R^3  \sin(\frac{\pi a}{R}) \rho _{DM}^{(c)} \pi \left\lbrace R^2 a \rho_{DM}^{(c)} \cos(\frac{\pi a}{R})-\frac{\pi (a+r_0)}{4} \right\rbrace  \\
&+& 8 R^2 a^2 \pi^3 \rho_{DM}^{(c)} \cos(\frac{\pi a}{R})+\pi^4 a r_0 \left\lbrace \ln(\frac{r_0}{R})-\ln(\frac{a}{R})\right\rbrace + 32 (\rho _{DM}^{(c)})^2 R^4 \left\lbrace R^2-a(a-r_0)\pi^2 \right\rbrace  \Big],
\end{eqnarray}
where 
\begin{eqnarray}
\Xi_1 &=& R^3 \rho_{DM}^{(c)} \sin(\frac{\pi a}{R}) -\frac{\pi^2 a \left[\ln(\frac{a}{R})-\ln(\frac{r_0}{R})-1\right]}{8},\\
\Xi_2 &=& R^3 \rho_{DM}^{(c)} \sin(\frac{\pi a}{R}) -\frac{\pi^2 a \left[\ln(\frac{a}{R})-\ln(\frac{r_0}{R})-2\right]}{8}.
\end{eqnarray}

It is interesting to see from Fig. 5 that there exits a domain of parameters where $\mathcal{I}_V>0$, implying a regular matter source which supports the BEC wormhole solution. In particular, for $r_0=1$, $\rho_{DM}^c=0.01$, and $R=9$, we find an interval such that the wormhole radius is situated in the interval $a \in [1,2.5]$, with $\mathcal{I}_V<0$. This really shows that near the wormhole throat exotic matter is needed to keep the wormhole throat open. Outside the wormhole region, for example in $a \in [2.6,7.7]$, we find $\mathcal{I}_V>0$. 
\begin{figure}[h!] 
\center
\includegraphics[width=0.70\textwidth]{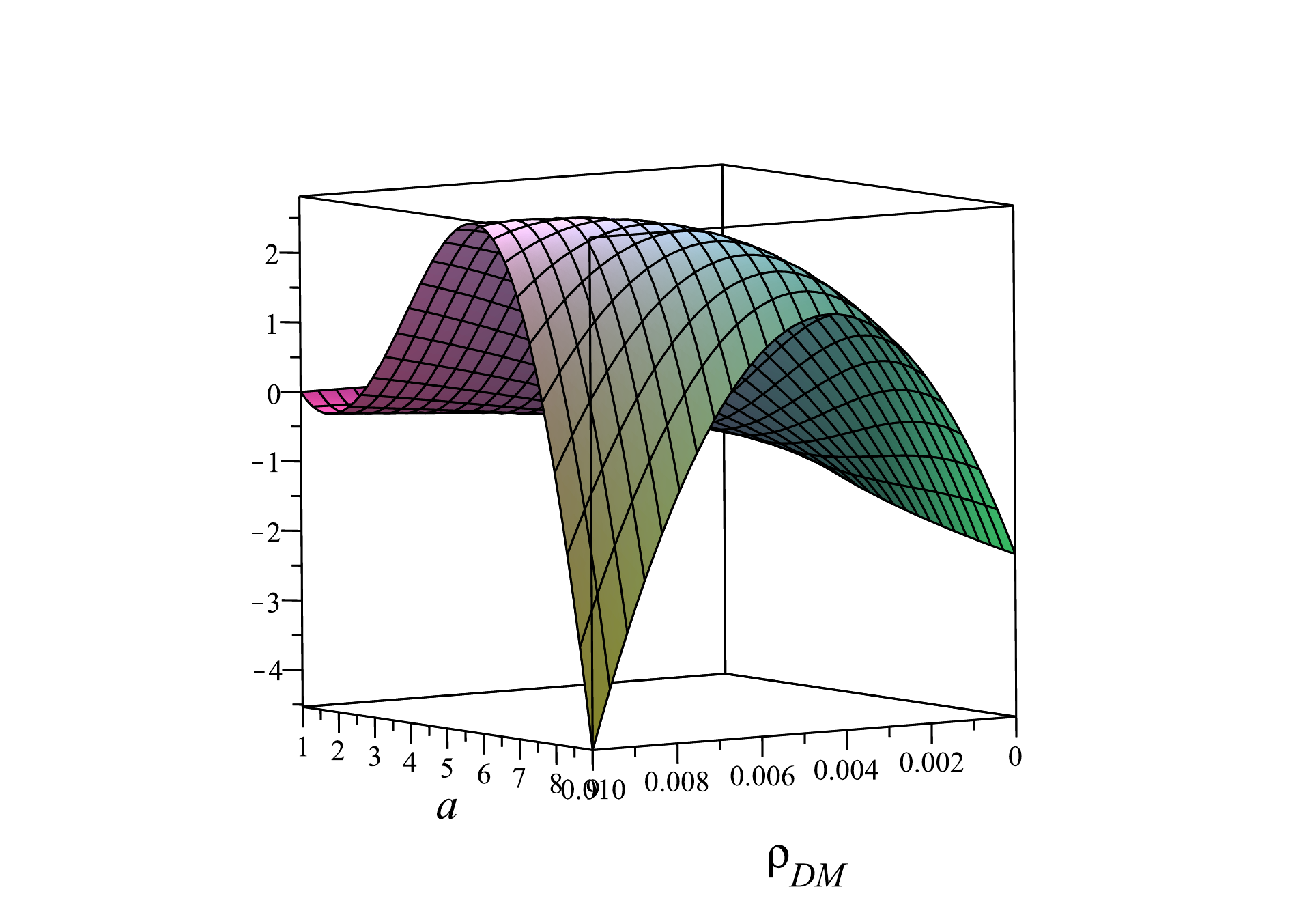}
\caption{{\protect\small \textit{The figure shows the variation of the
$I_{V}$ versus $a$ and $\rho_{DM}^c$. We have chosen $r_0=1$ and $R=9$, with $a \in [1,9]$ and $\rho_{DM}^c \in [0,0.01]$.  }}}
\end{figure}
\newpage
\section{A Kerr-like metric for rotating wormholes }
In this section we intend to use a rotating metric that guarantees the metric regularity of our wormhole solution. Following the arguments in Ref. \cite{azreg} we generalize our solution (27) by introducing a Kerr-like metric  which describes a rotating wormhole in BEC dark matter halo. The metric has the following form
\begin{eqnarray}\label{klwh}
ds^2&=&-\left(1-\frac{2 f}{\rho^2}\right)dt^2+\frac{\rho^2 A}{\Delta} \frac{dr^2}{1-b(r)/r}- \frac{4a f \sin^2\theta}{\rho^2}dt d\varphi+\rho^2 d\theta^2+\frac{\Sigma \sin^2\theta }{\rho^2}d\varphi^2
\end{eqnarray}
where 
\begin{eqnarray}
\rho^2&=& r^2+a^2 \cos^2\theta,\,\,\, 2f=r^2(1-A)\\
\Delta &=& r^2 A+a^2,\,\,\,\,\Sigma = (r^2+a^2)^2-a^2\Delta \sin^2\theta,\\
A&=&\exp\left[-\frac{8\,\rho _{DM}^{(c)} R^3}{\pi^2 r}\sin\left( \frac{ \pi r}{R}\right)\right]
\end{eqnarray}

\section{Lense-Thirring Precession of Test Gyroscope}
Due to rotation of a spacetime there are always frame dragging effects. 
To measure geodetic precession effects that are due to curvature and Lense-Thirring due to frame dragging of the earth NASA have launch the Gravity
Probe B (GP-B) in 2004 \cite{GPB1,GPB2}. Recently, the precession frequency effects of Kerr black hole \cite{LTkerr} and Kerr-like black holes in dark energy and dark matter have been discussed in the literature \cite{LTde,LTdm}. The precession frequency effects in a rotating traversable
wormhole also have been studied in the literature \cite{LTwh}.

In this section we study the Lense-Thirring precession frequency of a test gyroscope in a Kerr-like rotating wormholes. Consider an observer in stationary rotating spacetime of wormhole with timelike Killing vector field ${K}$. Then to remain at rest in this spacetime it has to move along an integral curve $\gamma(\tau)$ of Killing vector field ${K}$ with four velocity
\begin{eqnarray}
u=(-K^2)^{-1/2} K.
\end{eqnarray}\label{4velocity}
 The spin precession frequency of a gyroscope attached to this observer (with respect to a fixed star) coincides with the vorticity field associated with the Killing congruence and thus defined as \cite{book}
 \begin{equation}
 \tilde{\Omega}_{s}=\frac{1}{2\,K^{2}}\ast \left( \tilde{K}\wedge d\tilde{K}\right)
 \end{equation}
where * represents the Hodge star operator, $\wedge$ represent wedge product $\tilde{K}$ and $\tilde{\Omega}$ are one-forms of ${K}$ and $\Omega_{s}$. Note that here  $\ast \left( \tilde{K}\wedge d\tilde{K}\right)$ regarded as measure of absolute rotation. Now if we assume that the components of the metric tensor $g_{\mu\nu}$ of spacetime are independent of $t=x^0$ and the Killing vector ${K}=\partial_{0}$, then the precession frequency is called the LT precession frequency and is given by \cite{book}
\begin{equation}
\Omega_{LT}=\frac{1}{2}\frac{\varepsilon_{ijl}}{\sqrt{-g}}\left[g_{0i,j}\left(\partial_l-\frac{g_{0l}}{g_{00}}\partial_{0}\right)-\frac{g_{0i}}{g_{00}}g_{00,j}\partial_l\right],
\end{equation}\label{LTf}
where $\varepsilon_{ijl}$ is the Levi-Civita symbol, $g$ is the determinant of the metric $g_{\mu\nu}$ and "$,$" denotes the partial derivative. Using the metric components from \eqref{klwh} we get
\begin{equation}
\vec{\Omega} _{LT}=\frac{a}{B\Xi ^{3/2}\left( \Xi -2f\right) }\left[\left(2fB\sqrt{%
	\Delta }\cos \theta\right)  \hat{r}-\sin \theta \left( \Xi f_{,r}-2rf\right) \hat{%
	\theta}\right].
\end{equation}\label{LTexp}
with
\begin{equation}
B=\sqrt{\frac{re^{-\frac{C}{r}\sin kr}}{%
	\left\vert r-b(r)\right\vert	 }},\,\,\, C=\frac{8\ro^{(c)}_{DM}R^3}{\pi^2 } \quad \text{and} \quad k=\frac{\pi}{R}.
	\end{equation}
The magnitude of the LT precession frequency is 
\begin{equation}
\Omega _{LT}=\frac{a}{B\Xi ^{3/2}\left\vert \Xi -2f\right\vert }\left[
4f^{2}B^{2}\Delta \cos ^{2}\theta +\left( \Xi f_{,r}-2rf\right) ^{2}\sin
^{2}\theta \right] ^{1/2}.
\end{equation}

\begin{figure}[h!]
\includegraphics[width=0.45\textwidth]{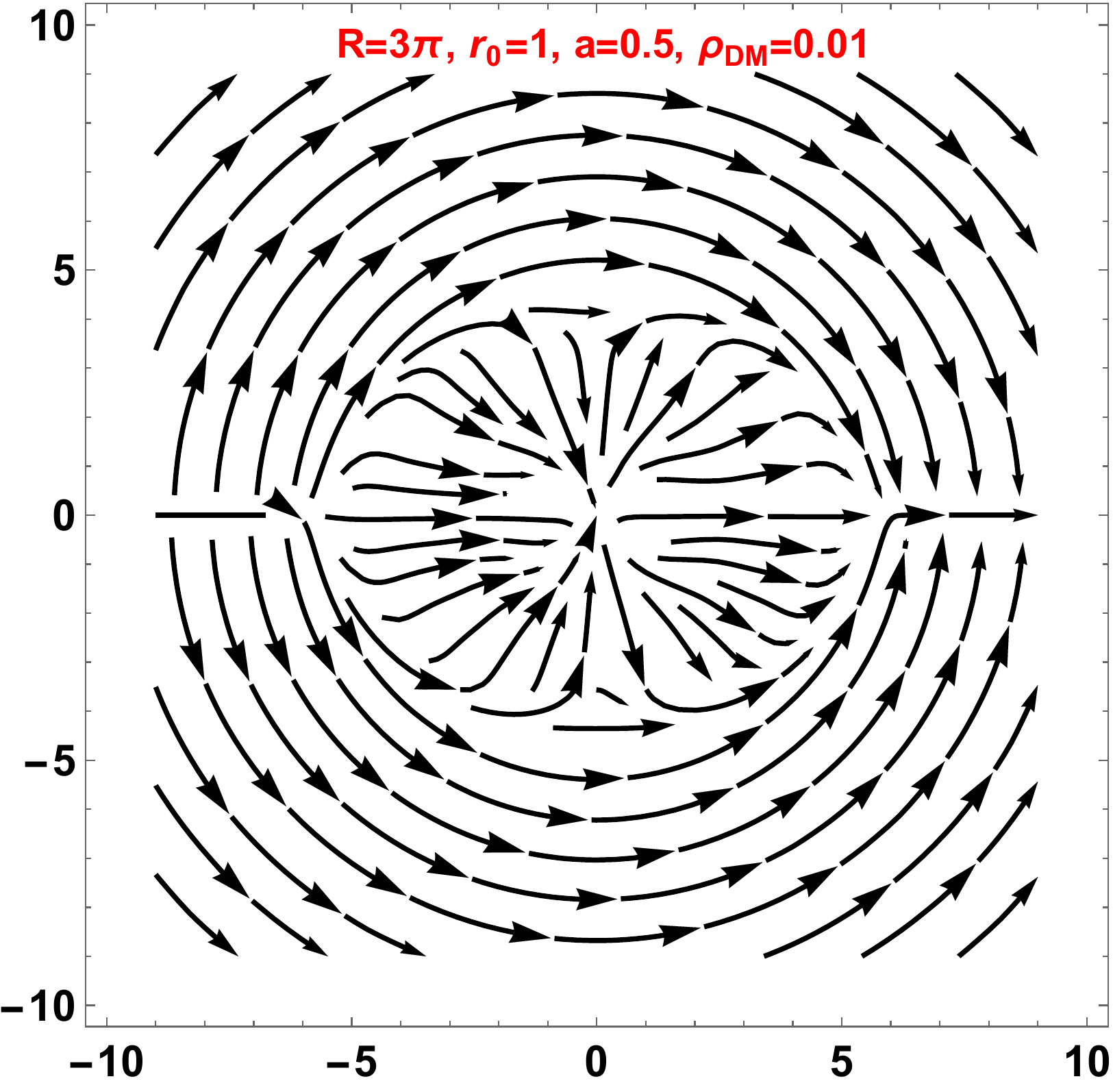}
\includegraphics[width=0.44\textwidth]{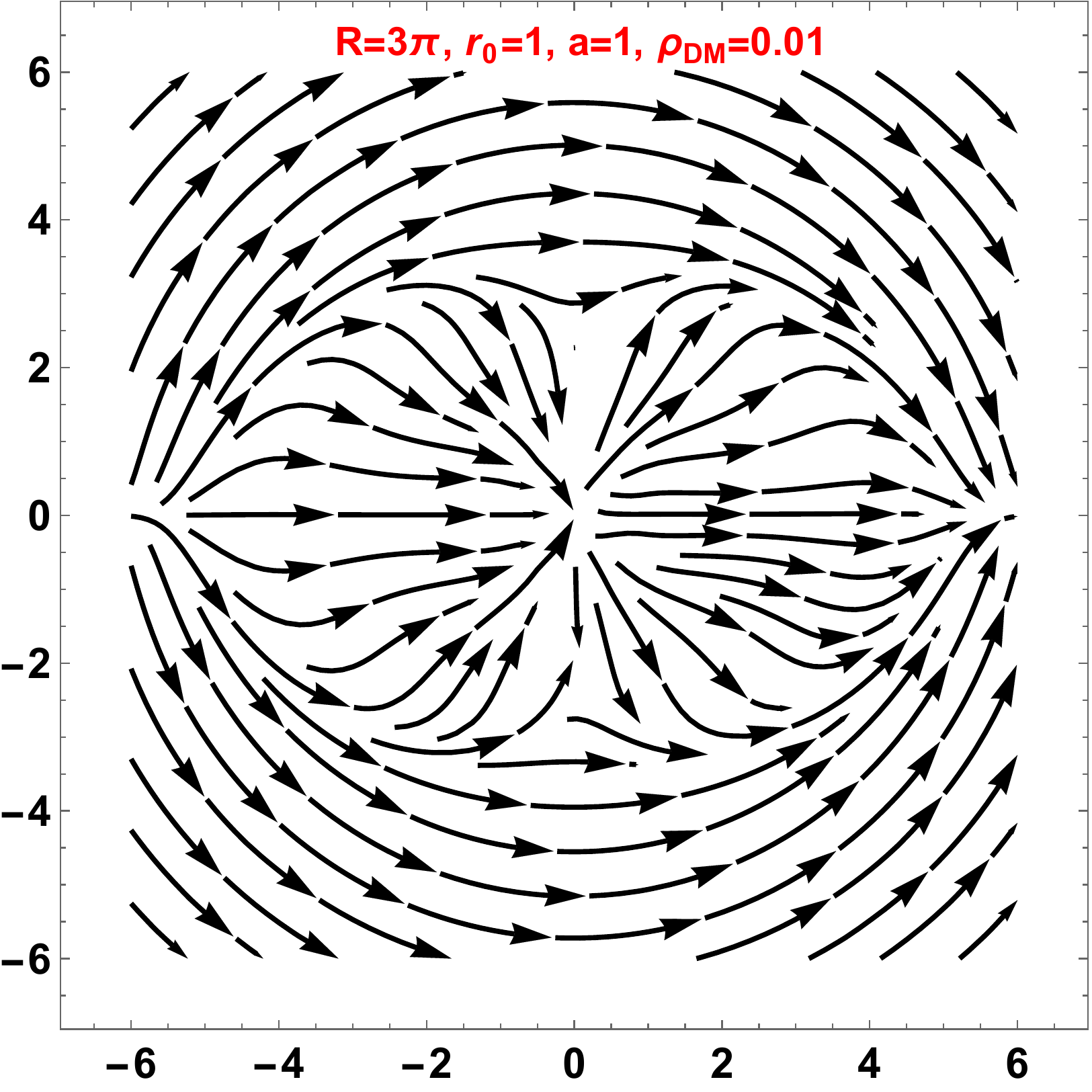}
\caption{{\protect\small \textit{ The vector field of the LT- precession frequency for a rotating wormhole is plotted. Due to the fact that the metric is regular, the vector field is defined inside the wormhole as well as outside the ergoshpere. }}}
\end{figure}

\section{Observational Aspect}
Recently, Zhou et al have studied the X-ray reflected spectrum of a thin accretion disk around the rotating Ellis wormhole \cite{Xraywh}. They have suggested that the wormholes may look like black holes and they have found some specific observational signatures by which it is possible to distinguish
rotating wormholes from Kerr black holes.
The rotating wormhole studied here in this manuscript may not be a compact object but if a particle moves in the spacetime of rotating wormhole it experiences some changes in periodic motion along spatial coordinates.

\begin{figure}[h!]
\includegraphics[width=0.47\textwidth]{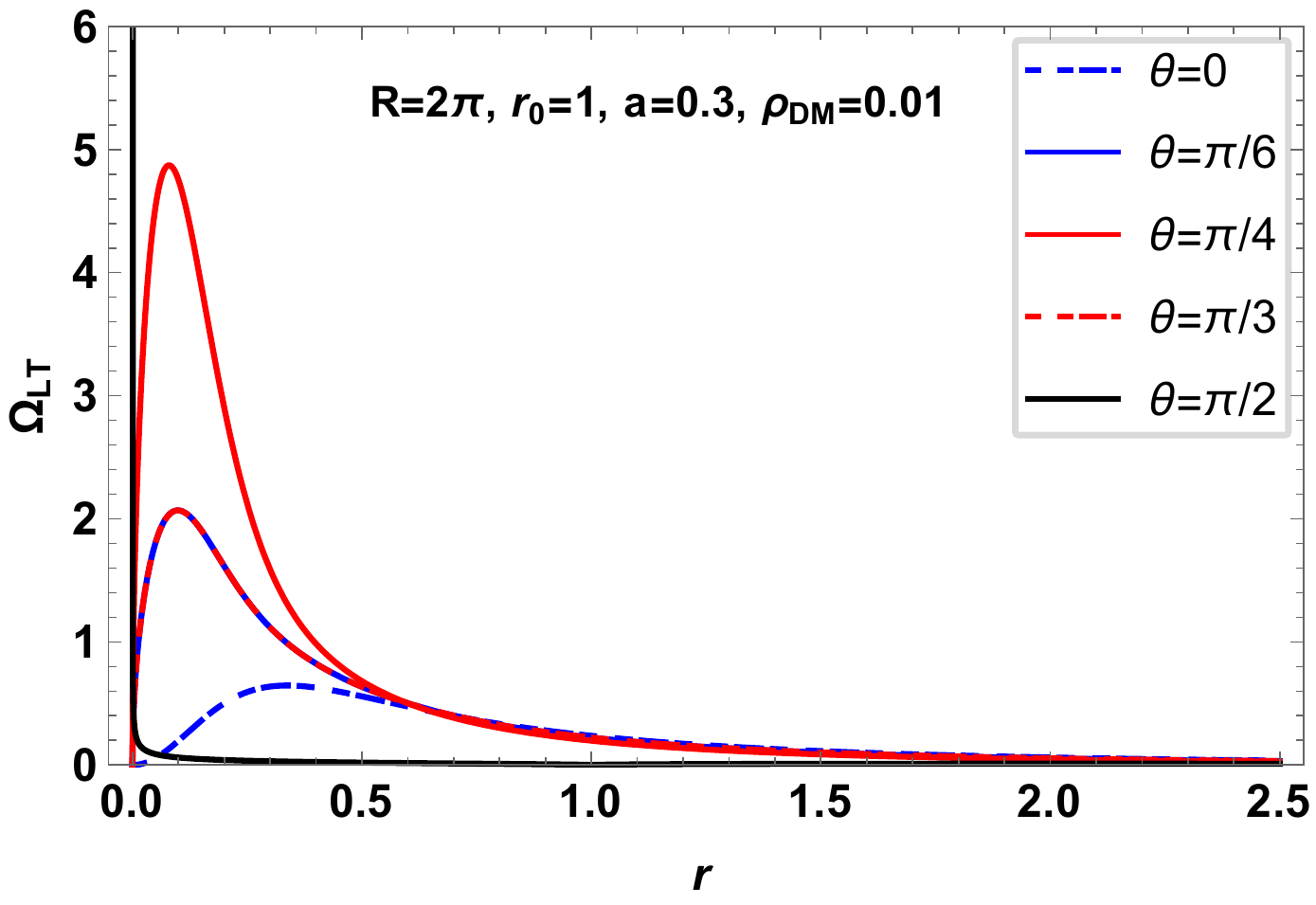}
\includegraphics[width=0.49\textwidth]{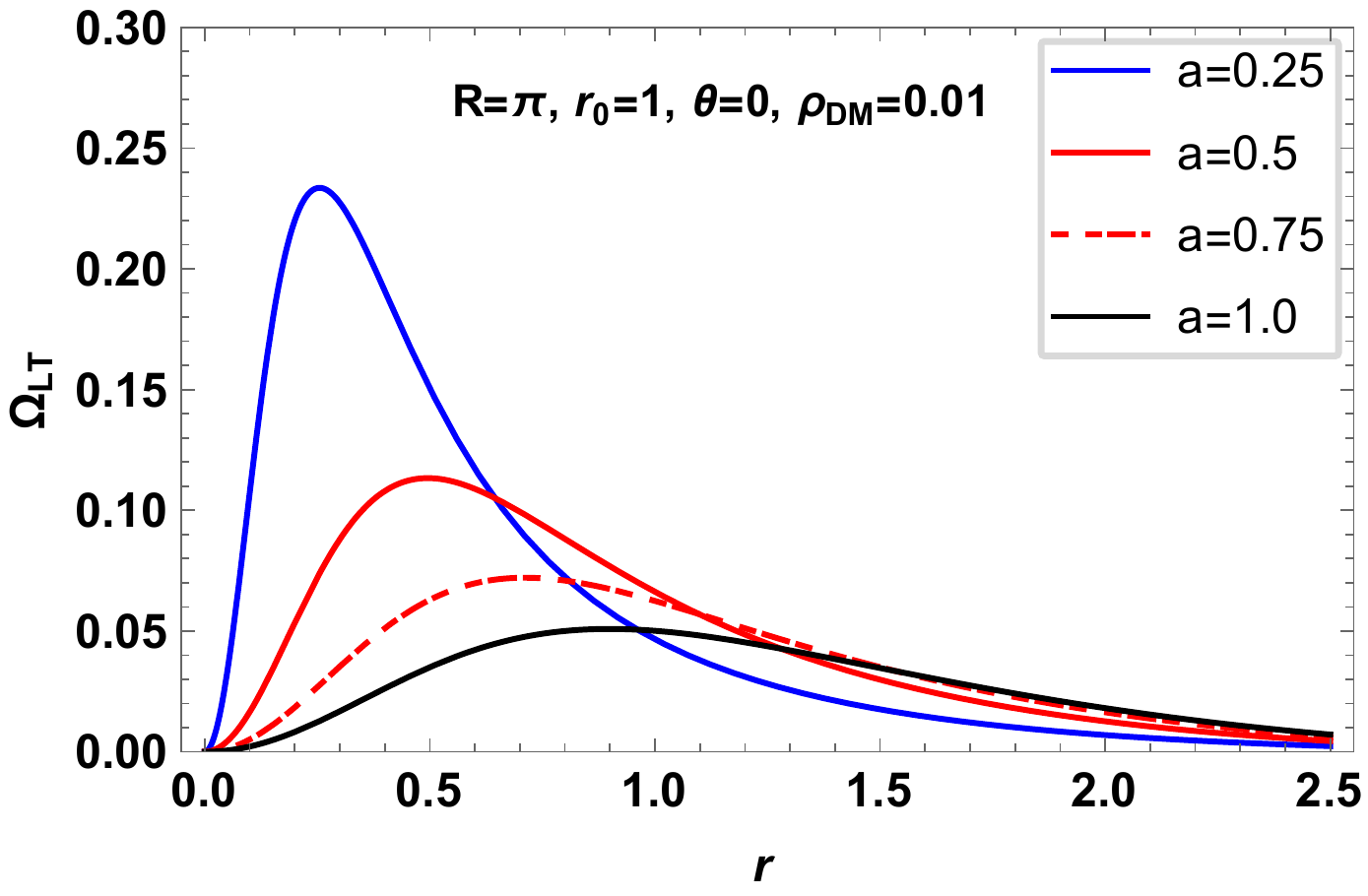}
\caption{{\protect\small \textit{ Here we plot $\Omega_{LT}$ 
 as a function of $r$ for different values of $\theta$. Left panel:Initially, with the increase of $r$ we observe that in all cases a particular peak value for  $\Omega_{LT}$ is obtained. Then $\Omega_{LT}$  decreases with the increase of $r$. Right panel: as the
observer approaches the wormhole $\Omega_{LT}$  
increases, then a particular peak value is obtained depending on the particular value $a$, and finally decreases with the decrease of $r$. }}}
\end{figure}

These changes in periodic motion along radial coordinate $r$, angular coordinate $\theta$ and $\phi$ are  characteristic by three frequencies of a test particle orbiting the wormhole are  the radial epicyclic frequency (REF) $\Omega_{r}$, the vertical epicyclic frequency (VEF) $\Omega_{\theta}$  and the Kepler frequency (KF) $\Omega_{\phi}$, respectively. The Kepler frequency of a particle is defined as \cite{3F} 
\begin{equation}
\Omega _{\phi }=\frac{d\phi }{dt}=\frac{-g_{t\phi }^{^{\prime }}\pm \sqrt{%
		g_{t\phi }^{^{\prime }}-g_{tt}^{^{\prime }}g_{\phi \phi }^{^{\prime }}}}{%
	g_{\phi \phi }^{^{\prime }}}\left\vert r=\text{cons., }\theta \text{%
	=cons.}\right. 
\end{equation}
where prime  denotes the derivative respect to "$r$". Here $+/-$ signs corresponds to direct/retrograde rotation, respectively. On the other hand, the radial epicyclic frequency is proportional to the second derivative of the effective potential and is associated with the quasi-periodic oscillations of X-ray binaries. This quantity is much smaller than the azimuthal frequency (see \cite{stella2} for details). Using the coefficients of metric tensor $g_{\mu\nu}$ from \eqref{klwh} we get
\begin{eqnarray}
\Omega _{\phi}=\frac{\sqrt{rA^{^{\prime }}}}{a\sqrt{rA^{^{\prime }}}%
	\pm \sqrt{2}r},
\end{eqnarray} 
where
\begin{eqnarray}
A^{^{\prime }}=\frac{C}{r^2}e^{-\frac{C}{r}\sin \left( kr\right) }\left[
-kr\cos \left( kr\right) +\sin \left( kr\right) \right].
\end{eqnarray}
The angular momentum of the particle is defined as 
\begin{eqnarray}
l=-\frac{g_{t\phi }+\Omega _{\phi }g_{\phi \phi }}{g_{tt}+\Omega _{\phi
	}g_{t\phi }},
\end{eqnarray}
The REC frequency of a particle at any fixed in the spacetime is defined as
\begin{widetext}
	\begin{eqnarray}
	\Omega _{r}^{2}=\frac{\left( g_{tt}+\Omega _{\phi }g_{t\phi }\right) ^{2}}{%
		2g_{rr}}\left[ \partial _{r}^{2}\left( \frac{g_{\phi \phi }}{\Gamma }\right)
	+2l\partial _{r}^{2}\left( \frac{g_{t\phi }}{\Gamma }\right) +l^{2}\partial
	_{r}^{2}\left( \frac{g_{tt}}{\Gamma }\right) \right] _{\left\vert r=\text{%
			cons., }\theta \text{=cons.}\right. }
	\end{eqnarray}
	where
	\begin{eqnarray}
	\Gamma =g_{tt}g_{\phi \phi }-g_{t\phi }^{2}.
	\end{eqnarray}
\begin{eqnarray}
\Omega _{r}^{2}=\frac{\left[ r-b\left( r\right) \right] \left[ -r\left\{
	a^{2}-r^{2}\left( 3A-2rA^{^{\prime }}\right) \mp 4ar\sqrt{2rA^{^{\prime }}}%
	\right\} A^{^{\prime }}+r^{2}\left( a^{2}+r^{2}A\right) A^{^{\prime \prime }}%
	\right] }{r^{3}A\left( \sqrt{2}r\pm a\sqrt{rA^{^{\prime }}}\right) ^{2}},
\end{eqnarray}

which gives

\begin{eqnarray}
\Omega _{r}=\frac{\sqrt{\left\vert r-b\left( r\right) \right\vert} \left[ -r\left\{
	a^{2}-r^{2}\left( 3A-2rA^{^{\prime }}\right) \mp 4ar\sqrt{2rA^{^{\prime }}}%
	\right\} A^{^{\prime }}+r^{2}\left( a^{2}+r^{2}A\right) A^{^{\prime \prime }}%
	\right] ^{1/2}}{\left( r^{3}A\right) ^{1/2}\left\vert \sqrt{2}r\pm a\sqrt{%
		rA^{^{\prime }}}\right\vert },
\end{eqnarray}
The VEC frequency is defined as
\begin{eqnarray}
\Omega _{\theta }^{2}=\frac{\left( g_{tt}+\Omega _{\phi }g_{t\phi }\right)
	^{2}}{2g_{\theta \theta }}\left[ \partial _{\theta }^{2}\left( \frac{g_{\phi
		\phi }}{\Gamma }\right) +2l\partial _{\theta }^{2}\left( \frac{g_{t\phi }}{%
	\Gamma }\right) +l^{2}\partial _{\theta }^{2}\left( \frac{g_{tt}}{\Gamma }%
\right) \right] _{\left\vert r=\text{cosntant, }\theta \text{=constant}%
	\right. }
\end{eqnarray}
\end{widetext}
which gives the frequency
\begin{eqnarray}
\Omega^{2} _{\theta }=\frac{2a\left( a\mp r\sqrt{2rA^{^{\prime }}}\right) \left(
	1-A\right) +r\left( r^{2}+a^{2}\right) A^{^{\prime }}}{r^{2}\left( \sqrt{2}%
	r\pm a\sqrt{rA^{^{\prime }}}\right) ^{2}},
\end{eqnarray}
and
\begin{eqnarray}
\Omega _{\theta }=\frac{\left[ 2a\left( a\mp r\sqrt{2rA^{^{\prime }}}\right)
	\left( 1-A\right) +r\left( r^{2}+a^{2}\right) A^{^{\prime }}\right] ^{1/2}}{%
	r\left\vert \sqrt{2}r\pm a\sqrt{rA^{^{\prime }}}\right\vert },
\end{eqnarray}

\begin{figure}[h!]
\includegraphics[width=0.45\textwidth]{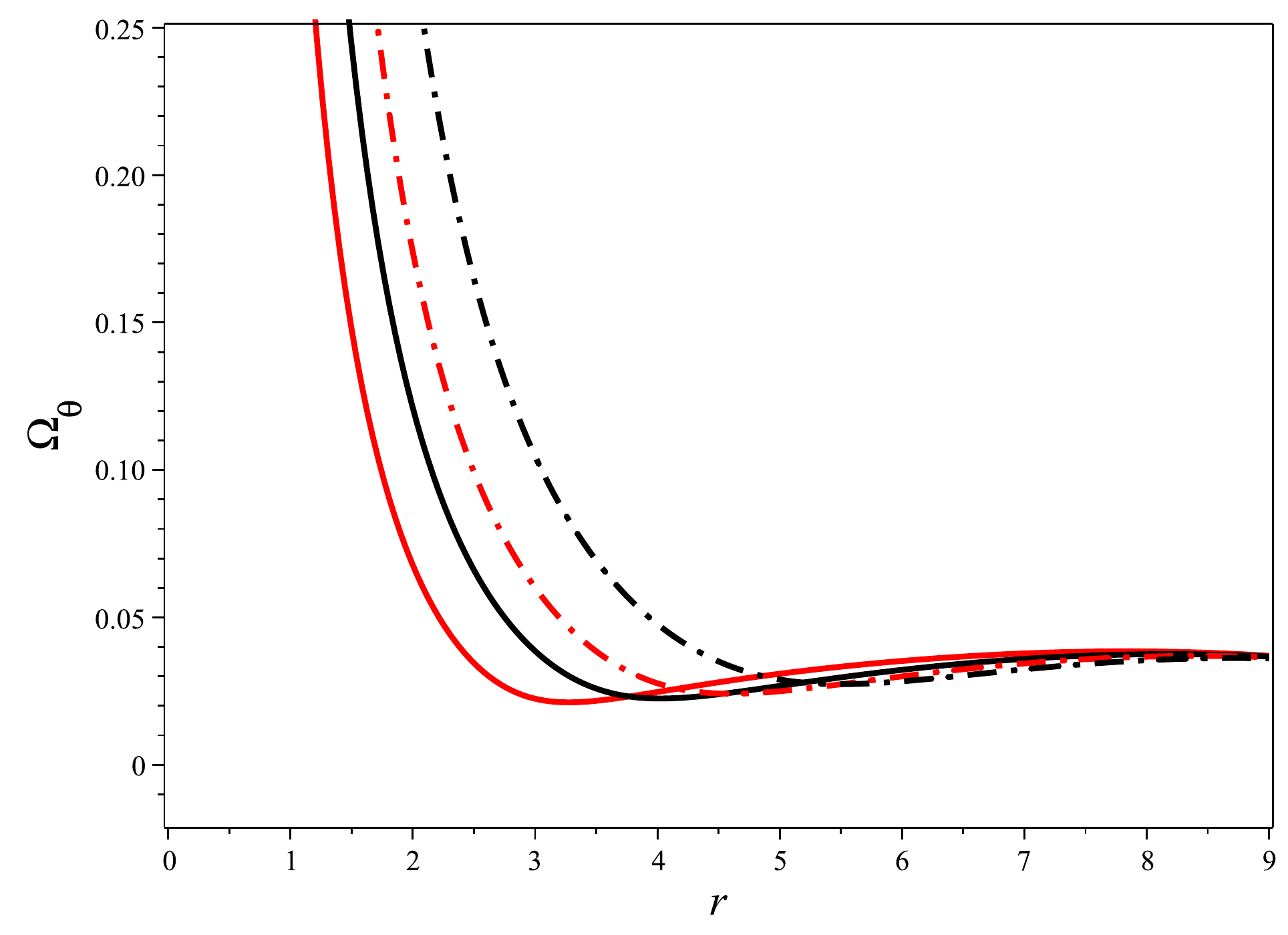}
	\includegraphics[width=0.45\textwidth]{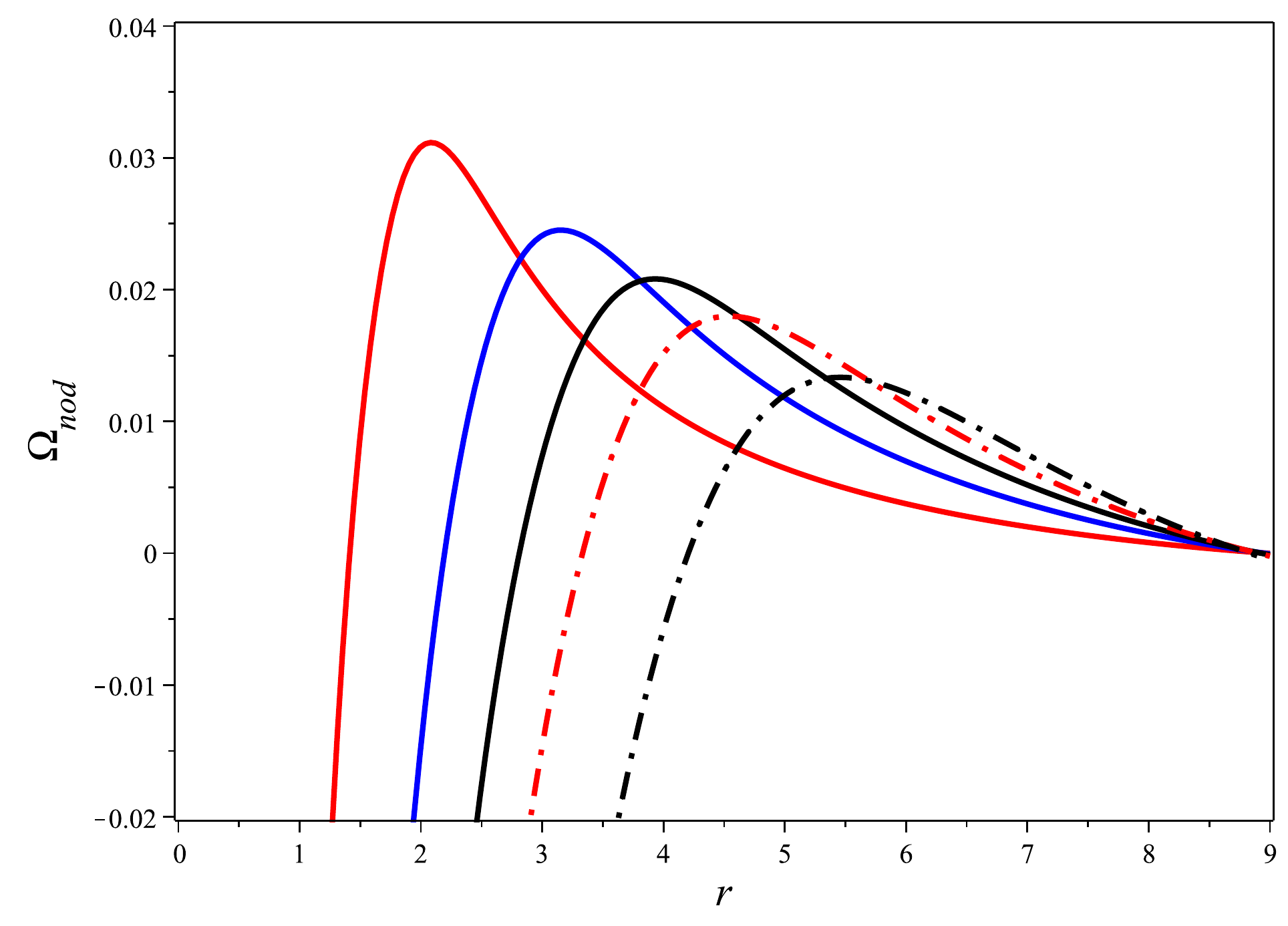}
	\caption{{\protect\small \textit{Left panel: The plane precession frequency $\Omega_{\theta}$  
				as a function of $r$ for different values of $a$ is plotted.  We have considered $R=9$, $\rho_{DM}=0.01$ and a dark matter halo mass $M_{DM}=10 {M}_\odot$ (=15 km). From left to right, $a=0.5$, $a=0.75$, $a=1$, and $1.5$,  respectively. Right panel:  It is shown the nodal precession frequency $\Omega_{\theta}$  
			as a function of $r$ for different values of $a$ is plotted. From left to right, $a=0.25$, $a=0.5$, $a=0.75$, $a=1$, and $1.5$,  respectively
	}}}
\end{figure}

Using the above relations, we can define the following two quantities 
\begin{equation}
\Omega_{\text{nod}}=\Omega _{\phi }-\Omega _{\theta },
\end{equation}
and 
\begin{equation}
\Omega_{\text{per}}=\Omega _{\phi }-\Omega _{r }.
\end{equation}

Where $\Omega_{\text{nod}}$ measures the orbital plane precession and is usually known as the nodal precession frequency(or Lense-Thirring precession frequency), on the other hand $\Omega_{\text{per}}$ measures the precession of the orbit and is known as the periastron precession frequency. Finally we point out that negative values of $\Omega_{nod}$, can be interpreted as a reversion of the precession direction.  
We provide a detailed analyses of our results in Fig. 9, where we highlight the observational aspects by calculating the impact of the BEC dark matter effect on $\protect\nu_{\text{nod}}$ which we can identify  with
HF QPOs.  To convert $\Omega_{nod}$ to $\nu_{nod}$, one can use  the relation $ \nu_{nod}$[k Hz]=$\Omega_{nod}$ [km$^{-1}$] $ \frac{300}{2 \pi } \times \frac{10 {M}_\odot}{M}$. From Fig. 9, we see that with the increase of $a$, and $r$ the  position of the peak shifts to the right, in other words, all frequencies become smaller and smaller with the increase of $r$. Interestingly, we see that the obtained frequencies belongs to the interval of typical QPOs  \cite{stella1,stella2,xray2,Revnew}. As of today, however, the correct explanation behind the QPO effect is not well understood, although QPOs are linked with the relativistic precession
of the accretion disk near black holes/neutron stars. In this direction, our result are interesting and may have astrophysical relevance. In particular here we have shown that the QPOs can be also linked with the relativistic precession of the rotating BEC wormhole. For example, by using these results one can distinguish BEC wormholes from black holes in the presence of BEC matter. The relativistic precession effect may shed some light in future experiments as an indirect way of detecting dark matter.
\begin{figure}[h!]
		\includegraphics[width=0.47\textwidth]{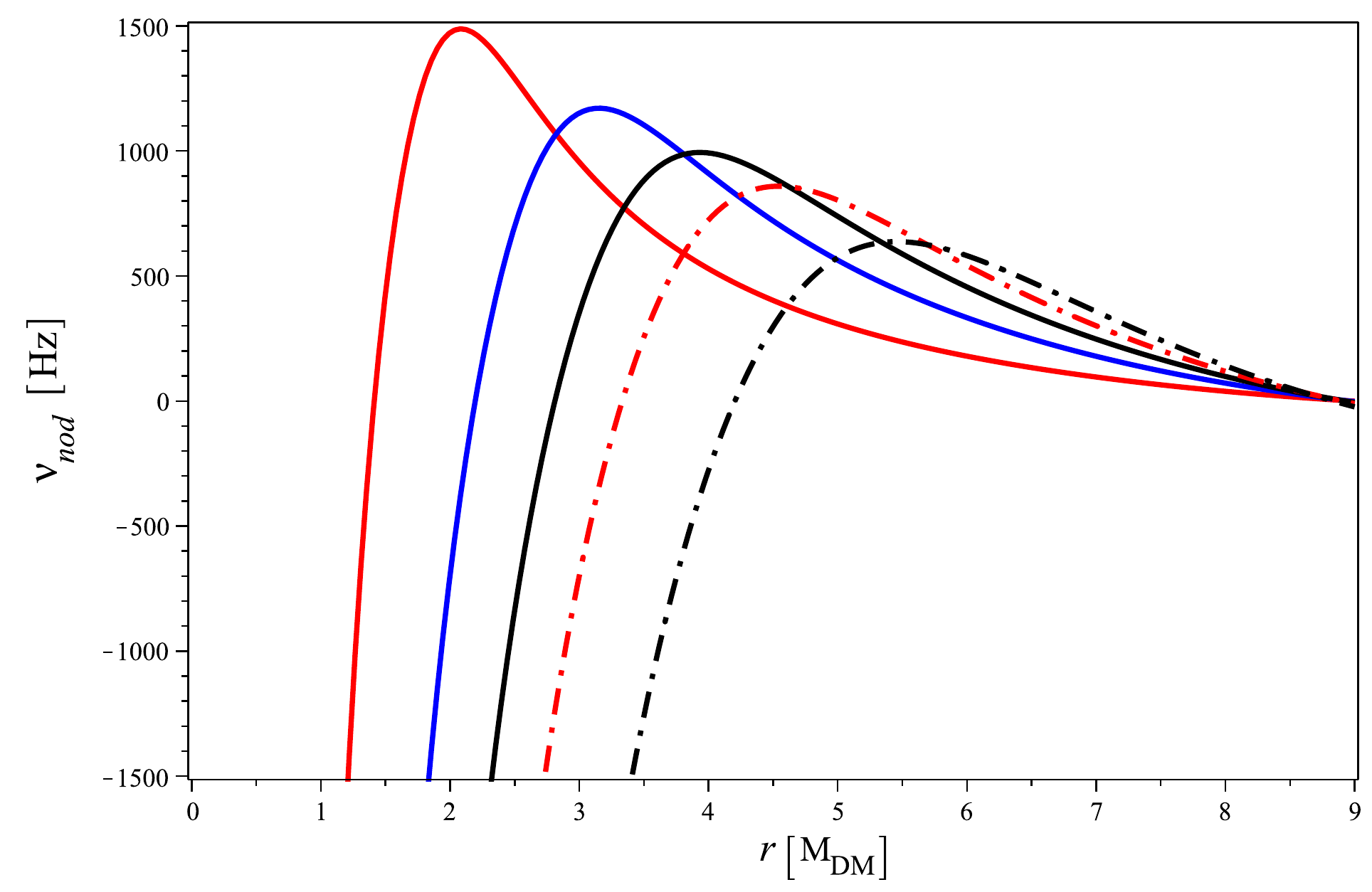}
	\caption{{\protect\small \textit{The nodal precession frequency $\nu_{nod}$  
			as a function of $r$ for different values of $a$ is plotted. We have considered $\rho_{DM}=0.01$ and the wormhole mass of $M=10 {M}_\odot$ (=15 km). From left to right, $a=0.25$, $a=0.5$, $a=0.75$, $a=1$, and $1.5$,  respectively. As we already pointed out the negative values of $\nu_{nod}$, can be interpreted as a
reversion of the precession direction.
	}}}
\end{figure}

\section{Conclusions}
In this paper, we have obtained a new wormhole solution supported by static and non-relativistic BEC.  More specifically, we have used the relation for the density profile of the BEC along with the rotation velocity to determine the wormhole red shift function and the shape function. To this end, we have calculated the radial and tangential pressures, respectively. 
It is shown that, for a specific choose of the central density of the condensates $\rho_{DM}^{c}$ and wormhole throat $r_0$ our wormhole solution satisfies the flare our condition at the wormhole throat. Furthermore we have checked the weak, strong and null energy condition at the wormhole throat. We find a domain of parameters such that $\rho \geq 0$, $\rho+2 \mathcal{P} \geq $ and $\rho+\mathcal{P}_r+2 \mathcal{P} \geq 0$,  together with the flare our condition are satisfied at the wormhole throat $r=r_0$. On the other hand, using the same parameters, we show that $\rho+\mathcal{P}_r\geq 0$ is violated by arbitrarily small quantities which implies that from a classical point of view,  energy conditions are violated at the wormhole throat. It is speculated that, from a quantum mechanical point of view  such small violations of the energy conditions can occur due to quantum fluctuations. Using the volume integral quantifier with a cut-off $a$ we have calculated the amount of exotic matter near the wormhole throat.

Introducing a Kerr-like metric for a rotating BEC wormhole we studied the LT effect. In particular, we have shown that the nodal frequencies lie within a range of observed typical QPOs, in other words QPOs can be linked with the relativistic precession of the rotating BEC wormhole. The interesting thing is that, the accretion disk is expected to change with time, therefore one can study the full evolution of QPO frequencies as the accretion disk, say, approaches the wormhole/black hole. A careful analysis of QPOs by future experiments can be potentially used to distinguish different objects, say  black holes from wormholes. The radius of BEC matter $R$ should be larger then the wormhole radius $r_0$, i.e. $R >r_0$. However, since the WH is supported by DM, which is quite plenty in the halo, there exists a possibility that the WH eats up all the halo DM and increase in size. In this case, $r_0=R$ ultimately occurs. It is known that a galaxy composed mainly of dark matter, however in a realistic situation, a galaxy also consists of a baryonic (normal) matter. In the preset paper, we considered only the effect of BEC dark matter, however it will be interesting to add the baryonic matter effect with a total mass $M=M_{DM}+M_{B}$. The effect of baryonic matter are expected to tiny modify our results, and we plan to study such a problem in the near future.\\

\end{document}